\documentclass[preprint,12pt,authoryear]{elsarticle}
\usepackage[round]{natbib} 

\usepackage[margin=2.5cm]{geometry}
\usepackage{enumitem}
\newlist{inlinelist}{enumerate*}{1}
\setlist*[inlinelist,1]{%
  label=(\roman*),
}
\usepackage{soul}
\usepackage{hhline}
\usepackage{subfig}
\usepackage{multirow}
\usepackage{xcolor}
\usepackage{graphicx}
\usepackage{upgreek}
\usepackage{amssymb}
\usepackage{amsfonts,amsthm,bm,amsmath} 
\usepackage{appendix}
\usepackage{times}
\usepackage{caption}
\usepackage{subfig}
\newcommand{\psubref}[1]{\protect\subref{#1}}
\usepackage{multirow}
\usepackage{xcolor}
\usepackage[linesnumbered,ruled,vlined]{algorithm2e}
\usepackage{tablefootnote}

\SetCommentSty{mycommfont}

\SetKwInput{KwInput}{Input}                
\SetKwInput{KwOutput}{Output}              

\usepackage[colorlinks]{hyperref} 
\hypersetup{ 
    colorlinks=true,       
    linkcolor=red,          
    citecolor=blue,        
    filecolor=magenta,      
    urlcolor=cyan           
}

\newcommand{\fref}[1]{Fig.~\ref{#1}}

\newcommand{\eref}[1]{Eq.~(\ref{#1})}

\newcommand{\sref}[1]{Section~\ref{#1}}
\newcommand{\tref}[1]{Table~\ref{#1}}

\begin{document}

\begin{frontmatter}

\title{Material-Response-Informed DeepONet and its Application to Polycrystal Stress-strain Prediction in Crystal Plasticity}
\author[]{Junyan He$^{1}$\corref{mycorrespondingauthor}}
\cortext[mycorrespondingauthor]{Corresponding author}
\ead{jimmy.he@ansys.com}
\author[]{Deepankar Pal$^{1}$}
\author[]{Ali Najafi$^{1}$}
\author[]{Diab Abueidda$^{3}$}
\author[]{Seid Koric$^{2,3}$}
\author[]{Iwona Jasiuk$^2$}

\address{$^1$ Ansys Inc., Canonsburg, PA, USA \\
$^2$ Department of Mechanical Science and Engineering, University of Illinois at Urbana-Champaign, Urbana, IL, USA \\
$^3$ National Center for Supercomputing Applications, University of Illinois at Urbana-Champaign, Urbana, IL, USA \\
}

\begin{abstract}
Crystal plasticity (CP) simulations are a tool for understanding how microstructure morphology and texture affect mechanical properties and are an essential component of elucidating the structure-property relations. However, it can be computationally expensive. Hence, data-driven machine learning models have been applied to predict the mean-field response of a polycrystal representative volume element to reduce computation time. In this work, we proposed a novel Deep Operator Network (DeepONet) architecture for predicting microstructure stress-strain response. It employs a convolutional neural network in the trunk to encode the microstructure. To account for different material properties, boundary conditions, and loading, we proposed using single crystal stress-strain curves as inputs to the branch network, furnishing a material-response-informed DeepONet. Using four numerical examples, we demonstrate that the current DeepONet can be trained on a single material and loading and then generalized to new conditions via transfer learning. Results show that using single crystal responses as input outperforms a similar model using material properties as inputs and overcomes limitations with changing boundary conditions and temporal resolution. In all cases, the new model achieved a $R^2$ value of above 0.99, and over 95\% of predicted stresses have a relative error of $\le$ 5\%, indicating superior accuracy. With as few as 20 new data points and under 1min training time, the trained DeepONet can be fine-tuned to generate accurate predictions on different materials and loading. Once trained, the prediction speed is almost $1\times10^{4}$ times faster the CP simulations. The efficiency and high generalizability of our DeepONet render it a powerful data-driven surrogate model for CP simulations in multi-scale analyses.
\end{abstract}


\end{frontmatter}

\section{Introduction}
\label{sec:intro}
It is well-known that the microstructure of polycrystalline materials significantly influences the material response \citep{zhao2021selective,keller2011microstructural}. Recent advancements in modern additive manufacturing (AM) techniques have allowed accurate control of the manufacturing process parameters, and the microstructure can be tailored \citep{liu2022additive} to achieve desired mechanical properties. Effective adjustment of process parameters to optimize mechanical properties relies on a thorough understanding of the process-structure-property relations for the AM process. To elucidate how process parameters such as laser power and scan speed affect the microstructure, process simulation models were developed to obtain the thermal history during the print. Cellular automata were used to simulate the development of microstructure during solidification \citep{herriott2019multi,zhang2019modeling,zhang2023additive}. To understand how the microstructure affects mechanical properties and evolves in subsequent mechanical loading, prior research has relied on crystal plasticity (CP) models \citep{zhang2007microstructure,bridier2009crystal,joshi2022finite,he2021polycrystal,cheng2015crystal} to capture the local influence on the grain microstructures. However, CP simulations can be quite computationally expensive and even prohibitive for large models. Therefore, fast-Fourier-transform-based CP models have been utilized \citep{paramatmuni2019crystal,cocke2023implementation}, and adaptive remeshing schemes for complex polycrystals have been used in finite-element (FE) methods \citep{phung2021surface,phung2019voxel} to improve CP simulation speeds. In the meantime, outside of classical computational mechanics, the recent improvement of GPU-accelerated high-performance computing and machine learning (ML) techniques has led to significant strides in neural network (NN) applications in various fields, including plasticity simulations. Physics-informed NNs have been successfully applied to solve path-dependent plasticity problems \citep{he2023deep,abueidda2021meshless,niu2023modeling}. Data-driven NNs have been used to predict the elastic-plastic stress-strain curves \citep{liu2022mechanistically,he2023exploring,roy2023data,gorji2020potential,abueidda2021deep}, and the contours of the plastic strain during plastic loading \citep{HE2024107258}. Similarly, ML techniques have been applied to CP, such as to predict the stress distribution within the representative volume element (RVE) \citep{ibragimova2022convolutional,frankel2020prediction}, the mean stress-strain response of the polycrystal RVE \citep{bonatti2022cp,martinitz2023artificial,ali2019application,miyazawa2019prediction,bishara2023machine,sadeghpour2022data}, and the yield locus of CP model with different grain textures \citep{fuhg2022machine,nascimento2023machine}. However, most of these works rely on simple feed-forward neural networks (FNN) or convolutional neural networks, and they use the material property values, Euler angles, and texture of the microstructure as inputs. Furthermore, most models were trained for a single set of material properties.  

Recently, the Deep Operator Network (DeepONet) was pioneered by \cite{lu2021learning}. It is an architecture designed to learn the underlying operator of the governing equation but not a specific solution, making it more generalizable to parameterized input functions, boundary conditions, and varying materials. The classical DeepONet architecture is characterized by a branch and a trunk network. The branch network encodes the parameterized input variables/functions, and the trunk network encodes the underlying geometry. The outputs of the branch and trunk were then combined via a dot product, producing the final NN output. In its original form, the branch and trunk networks are FNNs \citep{lu2021learning}. In contrast, others have proposed different variations of it, such as using a convolutional neural network to extract image-based inputs \citep{wu2024capturing,HE2023116277,oommen2022learning}, a recurrent neural network to predict time-dependent fields \citep{HE2024107258,he2023multi}, and adding additional data communication links to improve performance \citep{wang2021learning}. Moreover, DeepONets have been employed to solve computational mechanics problems in heat transfer \citep{koric2023data}, elasticity \citep{goswami2022deep}, plasticity \citep{HE2023116277} as well as fracture \citep{goswami2022physics}. Previous studies have also explored transfer learning on DeepONets to generalize the predictions to unseen inputs \citep{goswami2022deep,xu2023transfer}.

With the proven track record of DeepONet and its excellent generalization capability, it is of high research interest to apply this novel NN architecture to polycrystals in CP simulations. Specifically, this current work seeks to develop a versatile and highly generalizable framework that can be applied to different material properties, microstructure textures, as well as loading conditions, and it should be data-efficient without requiring many CP simulations as training data. This current work also looks beyond using material properties as direct inputs to the NN and explores a way to input material response and boundary conditions in a more mechanically meaningful manner by utilizing the response of single crystals. Since CP simulations are widely used in multi-scale analyses to generate homogenized stress-strain curves and yield functions needed by macro-scale models, this work focuses only on predicting the mean-field stress-strain curves of RVEs instead of trying to predict the local stress distribution. This paper is organized as follows: \sref{sec:methods} provides an overview of the crystal plasticity model, single crystal responses, as well as the proposed neural network architecture. \sref{sec:results} presents and discusses the performance of the NN model with four different numerical examples. \sref{sec:conc} summarizes the outcomes and limitations and highlights future works.

\section{Methods}
\label{sec:methods}
\subsection{Crystal plasticity model}
\label{sec:CP}
In this work, a CP model is used to capture the response of polycrystal RVEs. We adopted a finite deformation formulation, where the multiplicative decomposition of the deformation gradient $\bm{F}$ is employed \citep{reina2014kinematic}:
\begin{equation}
    \bm{F} = \frac{ \partial \bm{u}}{\partial \bm{X}} + \bm{I} = \bm{F}^e \bm{F}^p,
\end{equation}
where $\bm{F}^e$ and $\bm{F}^p$ denote the elastic and plastic deformation gradients, respectively. Only plasticity due to dislocation slip is considered; hence the plastic velocity gradient $\bm{L}^p$ is defined from Schmid's law as \citep{clayton2010nonlinear}:
\begin{equation}
    \bm{L}^p = \dot{\bm{F}^p} (\bm{F}^p)^{-1} = \sum^{N_{ss}}_{\alpha=1} \dot{\gamma}^\alpha \bm{P}^\alpha,
\end{equation}
where $N_{ss}$ and $\dot{\gamma}$ denote the number of slip systems and slip rate. For face-centered cubic crystals, the slip rate can be expressed as:
\begin{equation}
    \dot{\gamma}^\alpha = \dot{\gamma}_0 \,  {\rm{exp}}\left\{ \frac{-\Delta F }{ k_B T } \left[ 1 - \left(\frac{ |\tau^\alpha| - g_a }{g^\alpha - g_a }\right)^{p} \right]^{q} \right\} {\rm{sgn}}( 1 + {\rm{sgn}}( \tau^\alpha ) ),
\end{equation}
where $\dot{\gamma}_0$ is a reference slip rate, $k_B$ is the Boltzmann constant, $T$ is the temperature, $g^\alpha$ denotes the slip resistance of slip system $\alpha$, and sgn denotes the sign function. To model the material hardening behavior during plastic loading, the slip resistance can evolve according to a hardening law:
\begin{equation}
    \dot{g}^\alpha = q^{\alpha\beta} h^\beta |\dot{\gamma}^\beta|,
\end{equation}
where $q^{\alpha\beta}$ is a matrix relating the self- and latent-hardening of different slip systems and $h^\beta$ is the hardening modulus. Strain hardening behavior can saturate, and $g^\alpha$ is bounded between its initial value $g_0^\alpha$ and saturation value $g_{max}^\alpha$. Due to material nonlinearity, Newton-Raphson iterations are employed to iteratively solve for the current $\bm{F}^p$ at each element integration point. Once the plastic strain is known, the second Piola-Kirchhoff stress can be determined as:
\begin{equation}
    \bm{T}^{PK2} = \frac{1}{2} \mathbb{C} ( \bm{F}^{eT}  \bm{F}^e - \bm{I} ),
\end{equation}
where $\mathbb{C}$ denotes the fourth-order elasticity tensor. The elastic behavior of the grains is assumed to be anisotropic, modeled by a cubic elastic material model, and can be uniquely defined by three constants $C_{11}$, $C_{12}$, and $C_{44}$. The Cauchy stress $\bm{\sigma}$ is calculated from $\bm{T}^{PK2}$ as:
\begin{equation}
    \bm{\sigma} = \frac{1}{det(\bm{F})} \bm{F} \bm{T}^{PK2} \bm{F}^T.
\end{equation}

In this work, we limit ourselves to oligocrystalline polycrystals under plane-strain conditions. It is further assumed that all the grain orientations are in-plane and can be described by a single Euler angle in the range [-180\textdegree,180\textdegree].

\subsection{Single crystal behaviors and material-response-informed DeepONet}
\label{sec:NN}
By design, a CP model captures the influence of local microstructure (e.g., grain size, grain shape, and preferred grain orientations) on the elastic-plastic material behavior. Therefore, CP simulation results depend not only on the material properties of the single crystal and boundary conditions but also on the underlying microstructure, leading to significant variation in response. In this section, we aim to construct a novel DeepONet architecture to separate the influence of microstructure from the influence of material properties and boundary conditions and do so in a mechanically meaningful way. 

\begin{figure}[h!] 
    \centering
     \subfloat[]{
         \includegraphics[trim={0cm 0cm 0cm 0.cm},clip,width=0.32\textwidth]{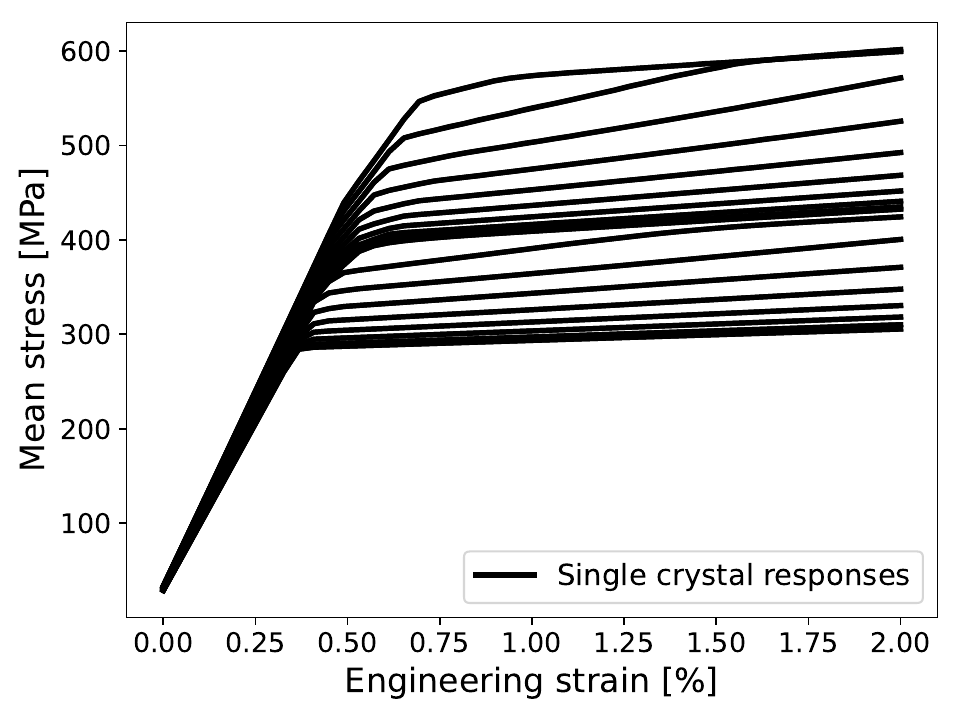}
         \label{basis}
     }
     \subfloat[]{
         \includegraphics[trim={0cm 0cm 0cm 0.cm},clip,width=0.32\textwidth]{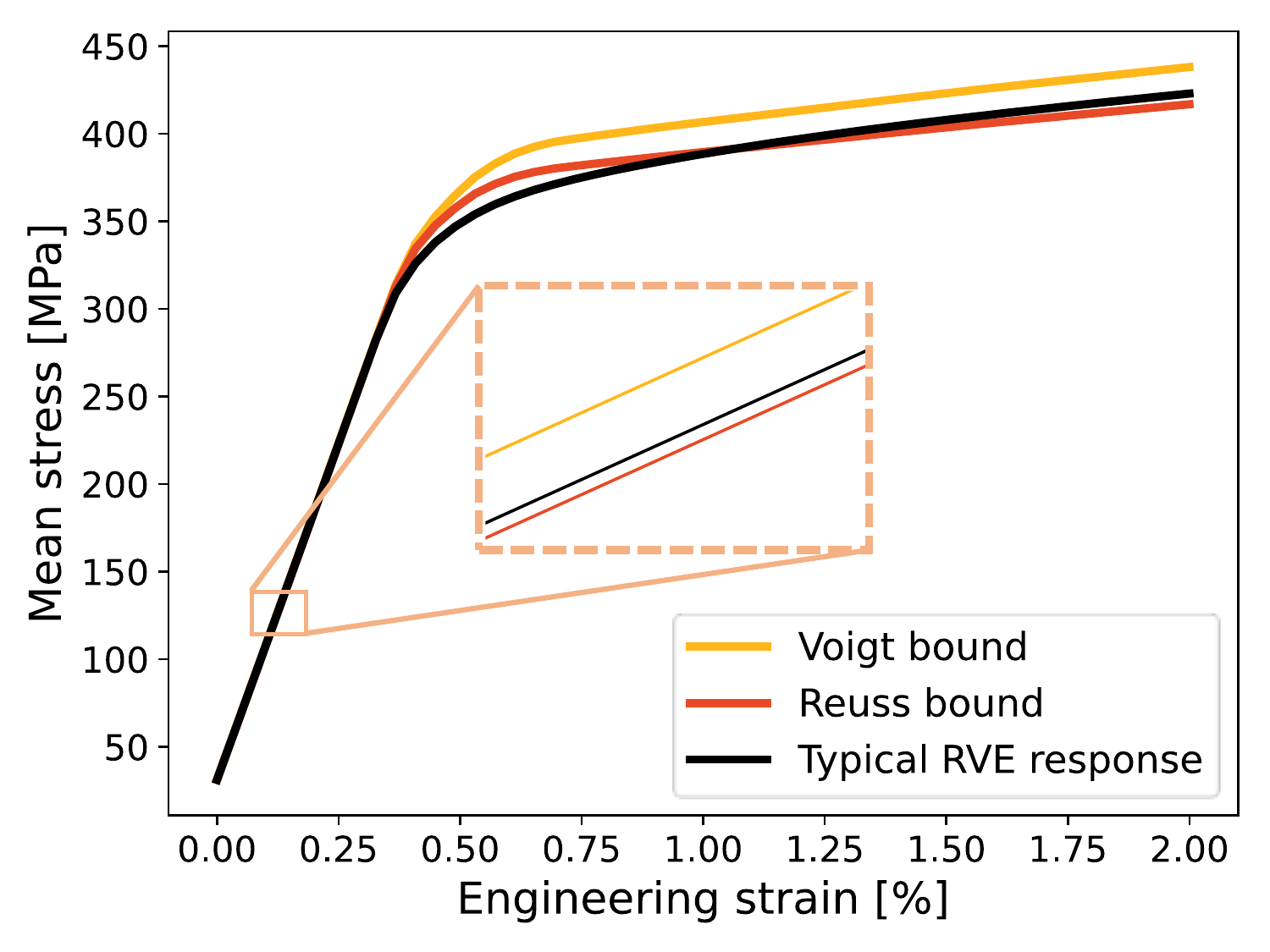}
         \label{basis_avg}
     }
     \subfloat[]{
         \includegraphics[trim={0cm 0cm 0cm 0.cm},clip,width=0.32\textwidth]{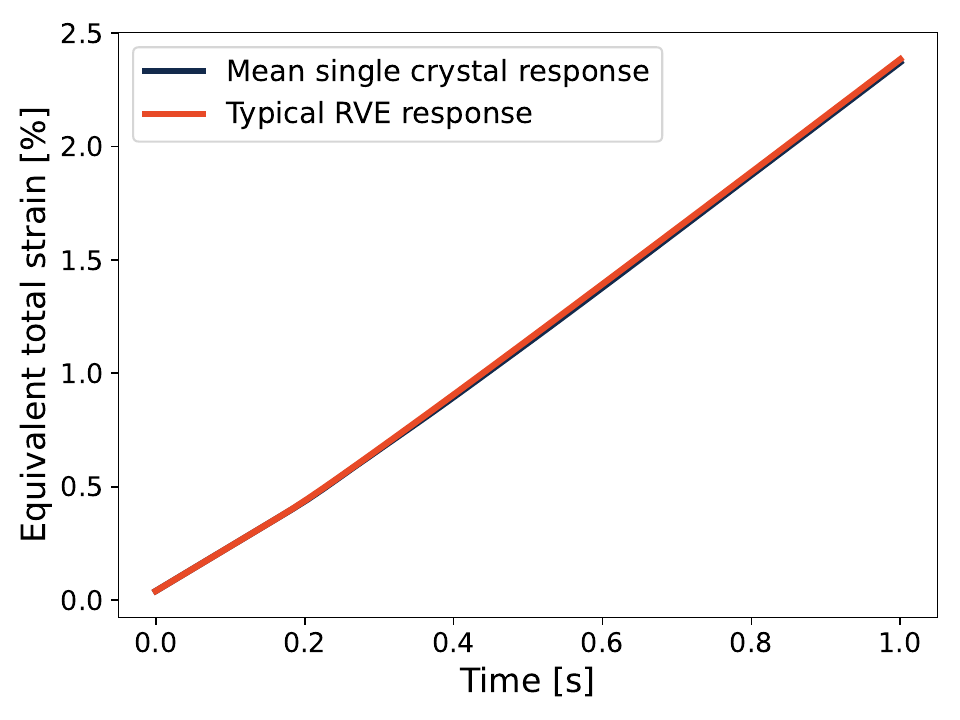}
         \label{avg_strain}
     }
    \caption{\psubref{basis} Single crystal responses under tension. \psubref{basis_avg} A typical polycrystal RVE response in tension compared to the Voigt and Reuss approximations computed from all single crystal responses. \psubref{avg_strain} Mean strain response of a typical polycrystal RVE and single crystals.}
    \label{single_cry_demo}
\end{figure}

To motivate the architecture of the DeepONet, first consider the elastic-plastic response of single crystals. Let a single crystal in the shape of a unit cube be discretized by \emph{one} hexahedral finite element. Each single crystal is a sample point in the full crystal orientation space. We employed 36 different single crystal sample points whose orientations are [4.86\textdegree, 9.73\textdegree, 14.59\textdegree, $\dots$, 175.13\textdegree], respectively. Note that only the positive half of the orientation space is sampled due to cubic material symmetry, although further simplification is possible using the crystal's four-fold symmetry. We also emphasize that this sampling is fixed and does not change with the microstructures of the different polycrystal RVEs considered in this work. Each single-crystal model is subject to the same boundary conditions and loading as the polycrystal RVEs of interest. In this work, we focus on the volume-averaged mean-field response, defined as:
\begin{equation}
    \hat{ \sigma } = \frac{1}{V_\Omega} \int_\Omega \sigma dV = \frac{1}{N^e} \sum_{i=1}^{N^e} \sigma^i,
\label{ms}
\end{equation}
where $\,\hat{}\,$ denotes a volume-averaged quantity, $V_\Omega$ is the volume of the domain, and $\sigma$ can be the scalar von-Mises stress or any stress tensor components. For a uniform mesh, the integral simplifies to a summation, and $N^e$ denotes the number of elements in the mesh. Taking tension as an example, the tensile stress-strain response of the single crystals is shown in \fref{basis}, while the global stress-strain response of a typical RVE is shown in \fref{basis_avg}. Two well-known rules of mixture, the \cite{voigt1889ueber} and \cite{reuss1929berechnung} models are shown in \fref{basis_avg} by considering the individual grains as different constituents in the mixture. Note that although the individual single crystal response differs drastically from that of the RVE, the approximations constructed from the two simple rules of mixture become close to the RVE response, hinting that a more complex rule of mixture may be determined to refine the approximation. This resemblance of the global RVE behavior with the two simple approximations is rooted in the fact that the average strain state between the RVE and the single crystals are similar (since identical boundary conditions are applied); see \fref{avg_strain}. 

\begin{figure}[h!] 
    \centering
         \includegraphics[width=0.95\textwidth]{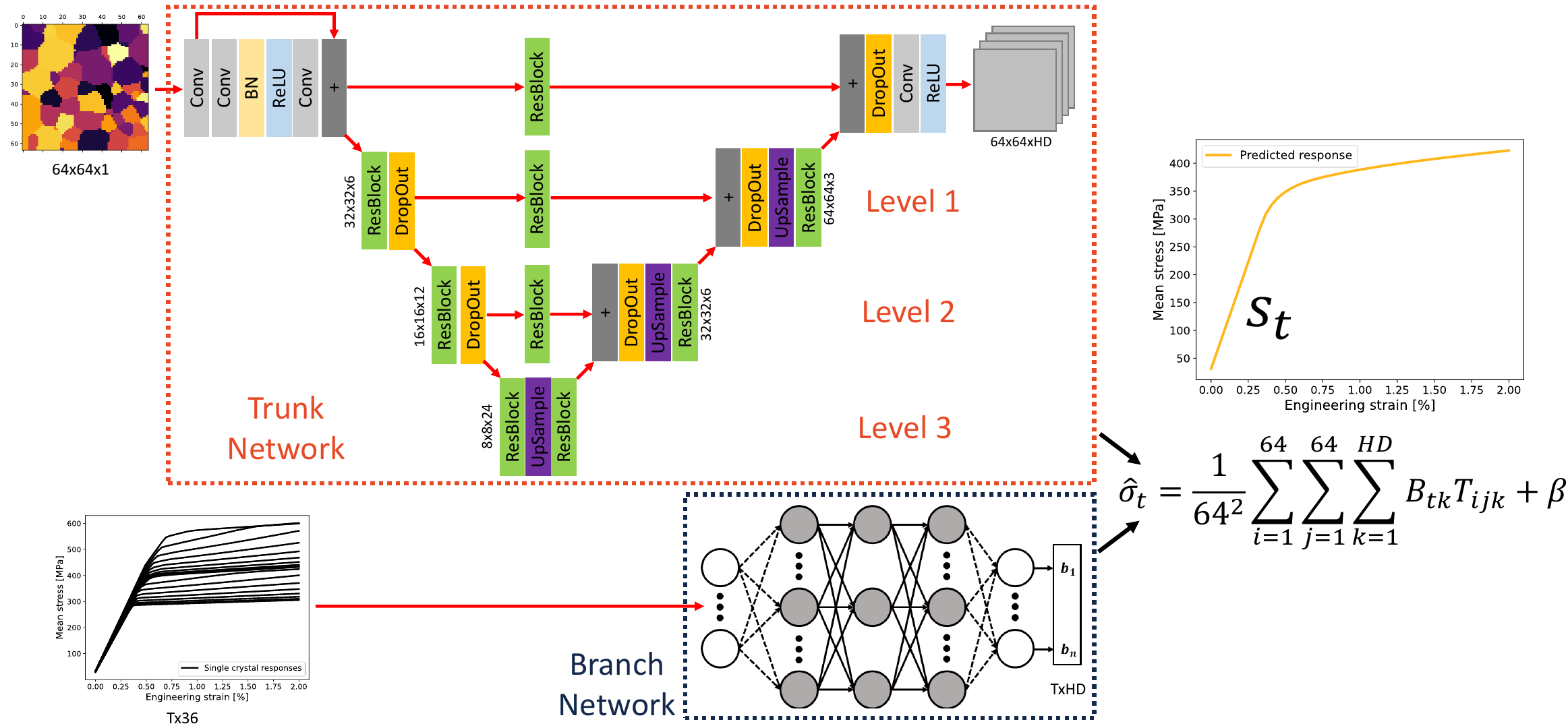}
    \caption{Schematic of the proposed DeepONet.}
    \label{schematic}
\end{figure}

From the observation above, we see that the single crystal responses (each computed from \emph{one} finite element) can be manipulated to resemble global RVE responses (computed from \emph{many} finite elements). From this perspective, we can view the single crystal stress-strain curves as "basis functions" (not necessarily orthogonal), and the global stress-strain curves of polycrystal RVEs under similar mean strain states can be approximated by a certain nonlinear rule of mixture to be determined by a NN. Most importantly, we emphasize that the single crystal responses contain information for both material properties \emph{and} boundary conditions. With that information implicitly represented in the form of single-crystal stress-strain curves, capturing the influence of local microstructure is necessary. The DeepONet architecture is well-suited for this problem structure. A typical DeepONet contains a trunk network that encodes the geometry of interest and a branch network that encodes various input functions that affect the network output. For the case of polycrystal RVE response prediction, the trunk network will encode the local microstructure measured by grain orientations to capture its influence, and the branch network will encode the single crystal responses (i.e., "basis functions" of the RVE response). Specifically, since microstructure can be described by a 2D array of grain orientations and the influence of the neighboring grains matters, a convolution neural network, the residual U-Net (ResUNet) \citep{diakogiannis2020resunet}, was used in the trunk. The trunk ResUNet takes 2D crystal orientation arrays in shape $[64\times64\times1]$ and generates its embedding of shape $[64\times64\times HD]$ where $HD$ denotes the hidden dimension of the DeepONet. The ResUNet used in this work has 3 encoding residual block levels and 24622 trainable parameters. We employed a simple feed-forward neural network (FNN) in the branch network to encode the 36 single-crystal responses in shape $[T\times36]$, where $T$ denotes the number of time steps in the single crystal stress-strain curves. The branch network then encodes the single crystal responses into shape $[T\times HD]$. $HD=16$ was used for all examples in this work. The FNN contains three layers of neurons: $[72 , 36 , 16]$ and has 5884 trainable parameters. A tensor-product operation combines the encoded outputs from the branch and trunk networks, followed by spatial averaging (similar to how the average of the stress at all finite elements yields the mean stress of the RVE, see \eref{ms}). This can be expressed as:
\begin{equation}
    \hat{\sigma}_t = \frac{1}{64^2} \sum^{64}_{i=1}\sum^{64}_{j=1} \sum^{HD}_{k=1} B_{tk} T_{ijk} + \beta,
\end{equation}
where $\hat{\sigma}_t$, $\bm{B}$, $\bm{T}$, and $\beta$ denote the predicted mean stress at time step $t$, branch output, trunk output, and a trainable scalar bias, respectively. A schematic of the proposed DeepONet is shown in \fref{schematic}.

All NNs were implemented in the DeepXDE framework \citep{lu2021deepxde} with a TensorFlow backend \citep{tensorflow2015-whitepaper}. The Python implementation of ResUNet was modified from the original code in \cite{resunet}. All models were trained for 80000 epochs with a batch size of 4 and a learning rate of $1\times 10^{-3}$. The Adam optimizer \citep{kingma2014adam} was used, and the scaled mean squared error (MSE) was used as the loss function, which is defined as:
\begin{equation}
    {\rm{MSE}} = \frac{ 1 }{ N } \sum^N_{i=1} (f_{FE} - f_{Pred})^2,
\end{equation}
where $N$, $f_{FE}$, and $f_{Pred}$ denote the number of data points, the FE-simulated values, and the NN-predicted values, respectively.

The proposed DeepONet can be trained on only \emph{one} set of material properties under \emph{one} loading condition to save training time and the required training data. This trained network is denoted as the baseline network. Transfer learning is used with minimal additional training data to extend the prediction to different material properties and/or different loading conditions. In addition to new RVE training data for the new material properties and boundary conditions, the corresponding single-crystal responses are also required. Since the convolutional layers are generally accepted as universal \citep{liu2021deep}, only the task-specific branch FNN is fine-tuned during transfer learning. The branch network is fine-tuned for 20000 epochs with a lower learning rate of $5\times 10^{-4}$. 

The most novel proposal in this work is the use of single crystal responses as a means to carry information on both the material properties and boundary conditions into the DeepONet instead of directly feeding the material property values. \cite{koric2023deep} previously investigated the use of DeepONet to capture stress contours for different material properties, where the material properties (i.e., Young's modulus, yield stress, and hardening modulus) are inputs to the branch network (see Figure 1 therein). This idea is adopted here to construct a baseline DeepONet model for performance comparison with our proposed architecture. To maintain a fair comparison, we only change the branch network in the baseline DeepONet, which is now an FNN taking 9 different scalar inputs and embedding them to an output shape of $[T\times HD]$ with a single layer of neurons. Doing so results in an FNN with 8000 trainable parameters, slightly more than that in the proposed DeepONet. The 9 input parameters consist of 8 material property parameters [$C_{11}$  , $C_{12}$ , $C_{44}$ , $h^\beta$ , $g_0^\alpha$ , $g_{max}^\alpha$ , $\dot{\gamma}_0$ , $\Delta F$] as well as 1 parameter on the loading condition, which is the magnitude of the applied strain. Since only a single scalar input is used to capture the different levels of applied strain, the baseline model cannot be extended toward different types of loading (e.g., from tension to shear). An identical training procedure was applied to train this baseline DeepONet. For easier reference, the baseline DeepONet taking material properties as inputs will be denoted as MP-DeepONet, while the proposed architecture taking single crystal responses will be called SC-DeepONet in subsequent discussions.

\subsection{Training data generation}
\label{sec:data_gen}
\begin{figure}[h!] 
    \centering
     \subfloat[]{
         \includegraphics[trim={0.5cm 1cm 0.5cm 1.cm},clip,width=0.235\textwidth]{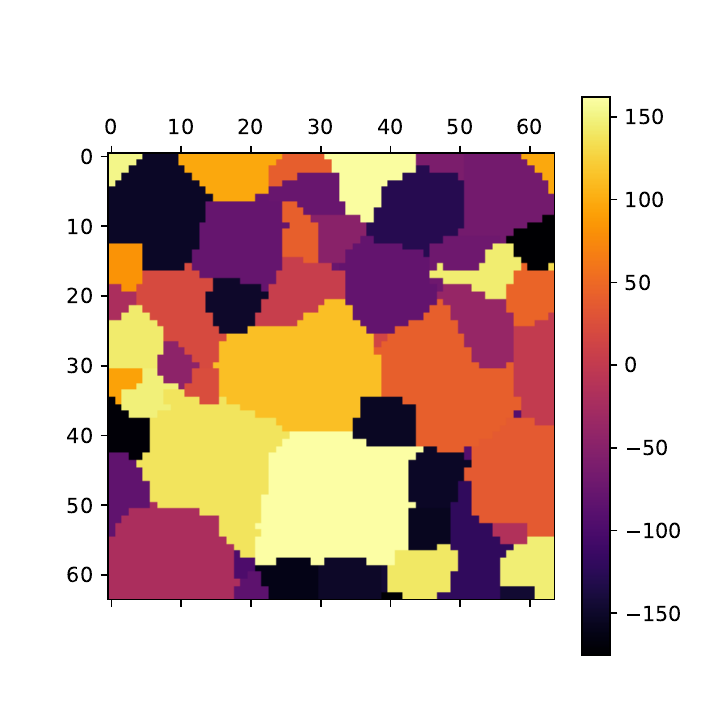}
         \label{rve0}
     }
     \subfloat[]{
         \includegraphics[trim={0.5cm 1cm 0.5cm 1.cm},clip,width=0.235\textwidth]{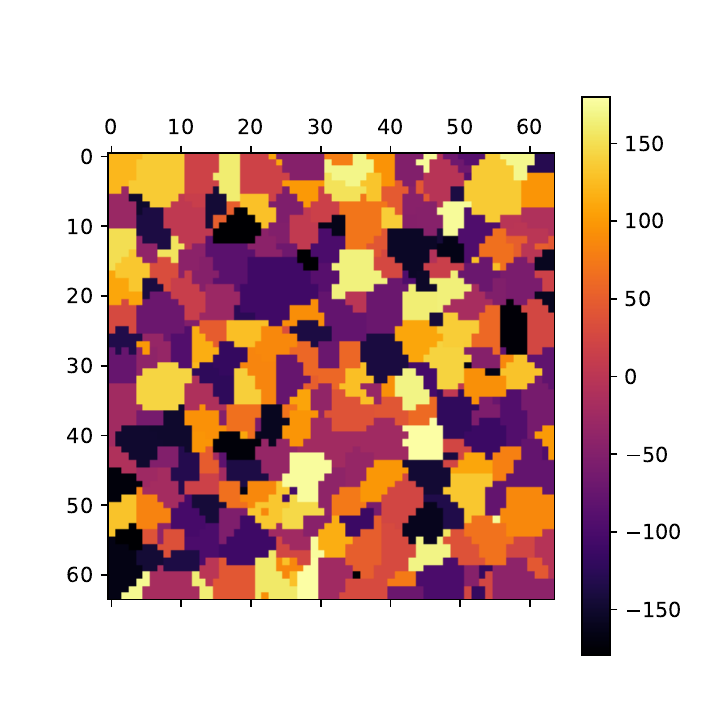}
         \label{rve1}
     }
     \subfloat[]{
         \includegraphics[trim={0.5cm 1cm 0.5cm 1.cm},clip,width=0.235\textwidth]{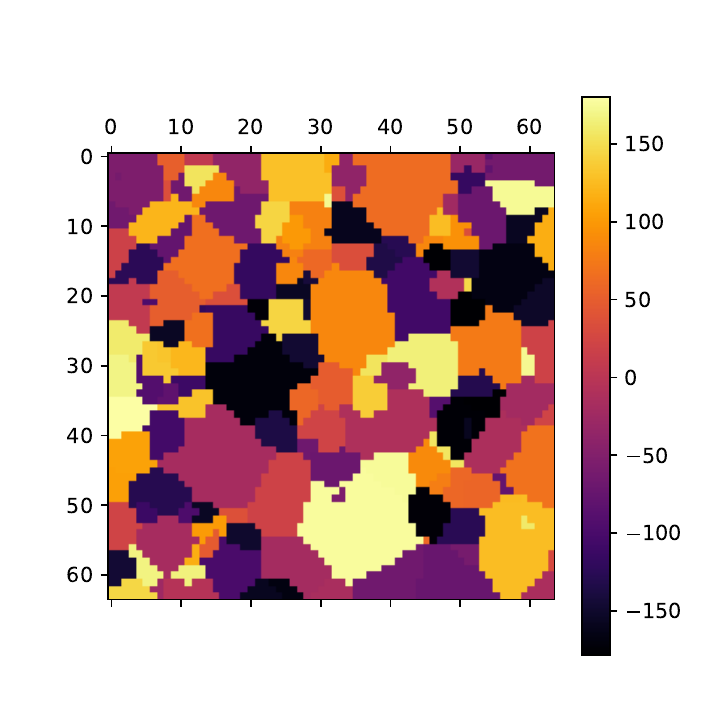}
         \label{rve2}
     }
     \subfloat[]{
         \includegraphics[trim={0.5cm 1cm 0.5cm 1.cm},clip,width=0.235\textwidth]{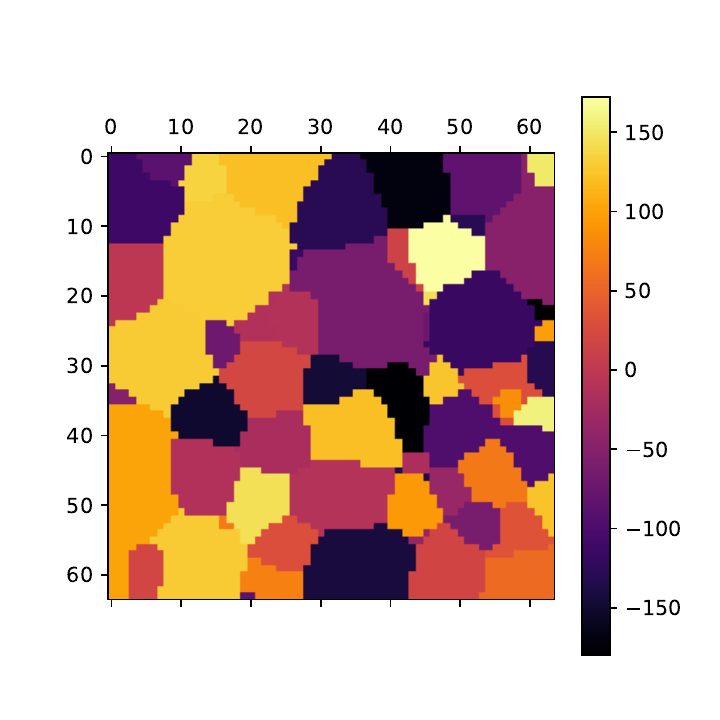}
         \label{rve3}
     }
    \caption{Four typical synthetic microstructures in the training data set, contour color is assigned based on Euler angle. }
    \label{micros}
\end{figure}

To generate diverse training data containing different grain morphologies, the synthetic microstructure generation tool DREAM.3D \citep{groeber2014dream} was used, and 1000 microstructures were generated with random Euler angles. Four polycrystal microstructures were selected and displayed in \fref{micros}. The microstructures were represented by a uniform mesh with $64\times64$ bi-linear hexahedral elements (Solid 185 in Ansys), and the edge length of the microstructures was set to 1 $mm$. The 3D finite elements were constrained in the thickness direction (out of the page, geometry was one element thick) to create a plane strain condition (the crystal plasticity model was implemented for 3D solid elements only; hence, the use of 2D plane-strain elements was not possible). The single crystal was modeled by a single hexahedral finite element (except for the shear case, where a $10\times10$ uniform mesh was used), and identical displacement boundary conditions were applied.
To test the extent of generalizability of the proposed network, especially towards different material properties and loading, two different materials and four different loading cases were considered. The first material was 2024-T3 aluminum alloy, and the second material was pure copper. The CP material parameters for aluminum were calibrated to experimental data provided in the work of \cite{esmaeili2015fatigue}. Selected CP material properties are shown in \tref{mat_props}.
\begin{table}[h]
\caption{\label{matprop} Material properties used in the crystal plasticity simulation\tablefootnote{Unit in MPa unless otherwise noted} }
\centering
\begin{tabular}{cccccccccc}
\hline
& $C_{11}$  & $C_{12}$ & $C_{44}$ & $h^\beta$ & $g_0^\alpha$ & $g_{max}^\alpha$ & $\dot{\gamma}_0$ [$s^{-1}$] & $\Delta F$ [J] \\
\hline
Al & 131038 & 80314 & 30942 & 1750 & 220 & 400 & 1.732$\times 10^{6}$ & 2.5$\times 10^{-19}$ \\
Cu & 166957 & 120900 & 75000 & 187 & 22 & 153 & 1.52$\times 10^{7}$ & 2.85$\times 10^{-19}$ \\
\hline
\end{tabular}
\label{mat_props}
\end{table}
Four different loading conditions and material combinations were considered in this work:
\begin{enumerate}
    \item Tension in the Y direction, 1\% engineering strain, aluminum
    \item Tension in the Y direction, 0.125\% engineering strain, copper
    \item Shear in the X-Y plane, 2\% engineering strain, aluminum
    \item Sinusoidal fully-reversible cyclic loading for 1.5 cycles, 0.125\% maximum strain, copper
\end{enumerate}

Ansys Mechanical 2023R2 was used to conduct the CP simulations on different microstructures, material properties, and loading conditions. An in-house python code was developed to convert the microstructure data in .vtk format (generated from DREAM.3D) to material assignment commands in Ansys parametric design language and pass the crystal properties and grain orientation to the solver. To account for the material behavior nonlinearity and path-dependent plasticity, all CP simulations were solved using a full Newton solver with automatic time stepping. The minimum and maximum step sizes used were $2\times10^{-3}s$ and $1\times10^{-2}s$, respectively. All external loads were applied as displacement boundary conditions and were ramped linearly over the step time (1s for cases 1-3, 3s for case 4). For load cases 1-3, the mean von-Mises stress $\sigma_{VM}$ of the RVE was stored in 50 uniform time steps, and it is defined as:
\begin{equation}
    \sigma_{VM} = \sqrt{ \frac{3}{2} \bm{S} : \bm{S} },
\end{equation}
where $\bm{S} = \bm{\sigma} - \frac{1}{3}tr(\bm{\sigma})\bm{I}$ is the deviatoric part of the stress tensor. For the cyclic loading case, the axial stress component $\sigma_y$ was stored in 240 time steps. It is best to scale the input and output data to suitable ranges for better NN training and accuracy. The input Euler angles were scaled linearly to the range $[0,1]$. At each time step, the RVE stress response was scaled using a min-max scaler with the data ranges determined from the corresponding single crystal responses at that time step.

\section{Results and discussion}
\label{sec:results}
All simulations were conducted with two high-end AMD EPYC 7763 Milan CPU cores. All NN training and inference were conducted using a single Nvidia A100 GPU card on Delta, an HPC cluster hosted at the National Center for Supercomputing Applications (NCSA). To evaluate the model performance in the test set, three quantitative metrics were used. They are the relative error, mean absolute error, and $R^2$ value:
\begin{equation}
\begin{aligned}
    {\rm{Relative \; error}} = \left| \frac{ f_{FE} - f_{Pred} }{ f_{FE} } \right | \times 100\%,\\
     {\rm{MAE}} = \frac{1}{N_T} \sum_{i=1}^{N_T} \left |  f_{FE} - f_{Pred} \right |,\\
    R^2 = 1 - \frac{ \sum_{i=1}^{N_T} \left( f_{FE} - f_{Pred} \right)^2 }{  \sum_{i=1}^{N_T} \left( f_{FE} - \Bar{f}_{FE} \right)^2 } ,
\end{aligned}
\end{equation}
where $f_{FE}$, $f_{Pred}$, $N_T$, $\Bar{f}_{FE}$ denote the finite element (FE) simulated field value, NN-predicted field value, number of test cases, and the mean value of the FE-simulated field values, respectively.

\subsection{Comparison between single crystal and RVE responses}
\label{sec:response_comp}
\begin{figure}[h!] 
    \centering
     \subfloat[]{
         \includegraphics[trim={0cm 0cm 0cm 0cm},clip,width=0.245\textwidth]{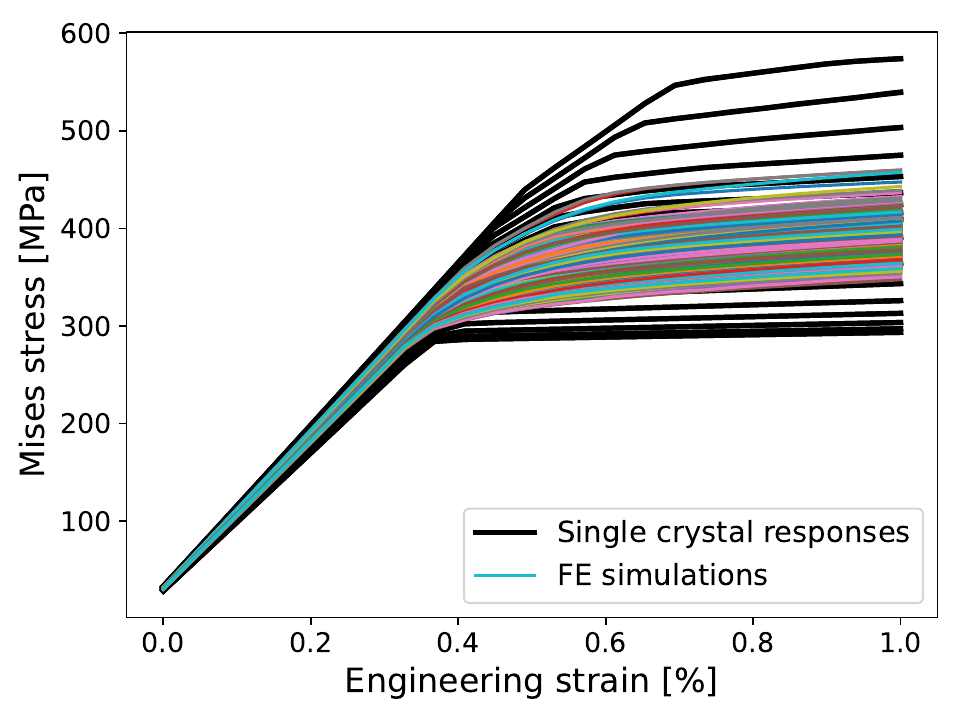}
         \label{c1}
     }
     \subfloat[]{
         \includegraphics[trim={0cm 0cm 0cm 0cm},clip,width=0.245\textwidth]{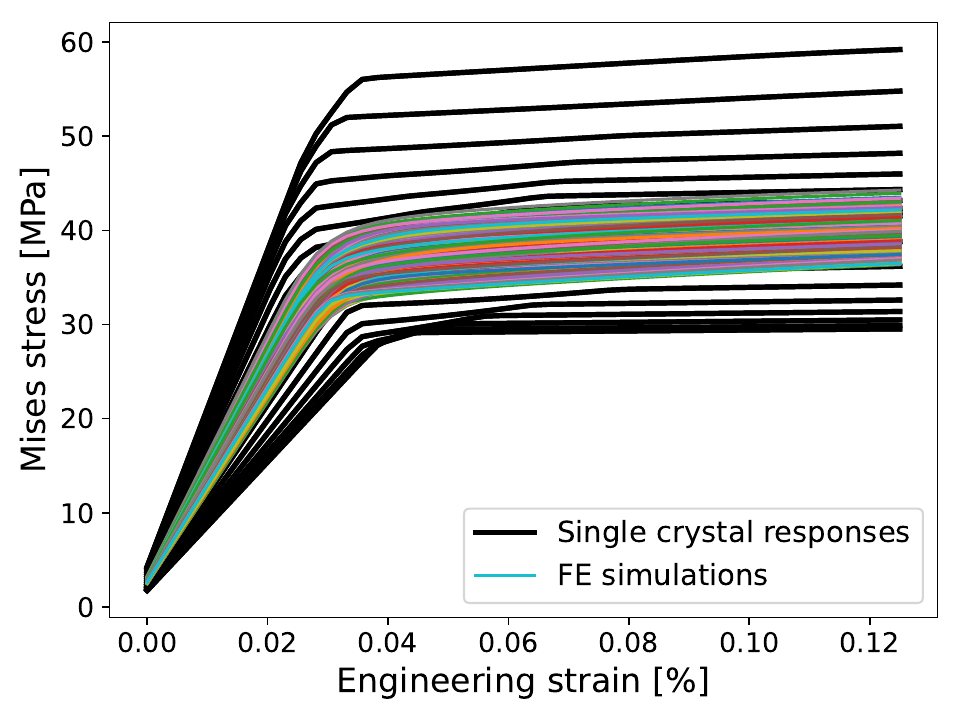}
         \label{c3}
     }
     \subfloat[]{
         \includegraphics[trim={0cm 0cm 0cm 0cm},clip,width=0.245\textwidth]{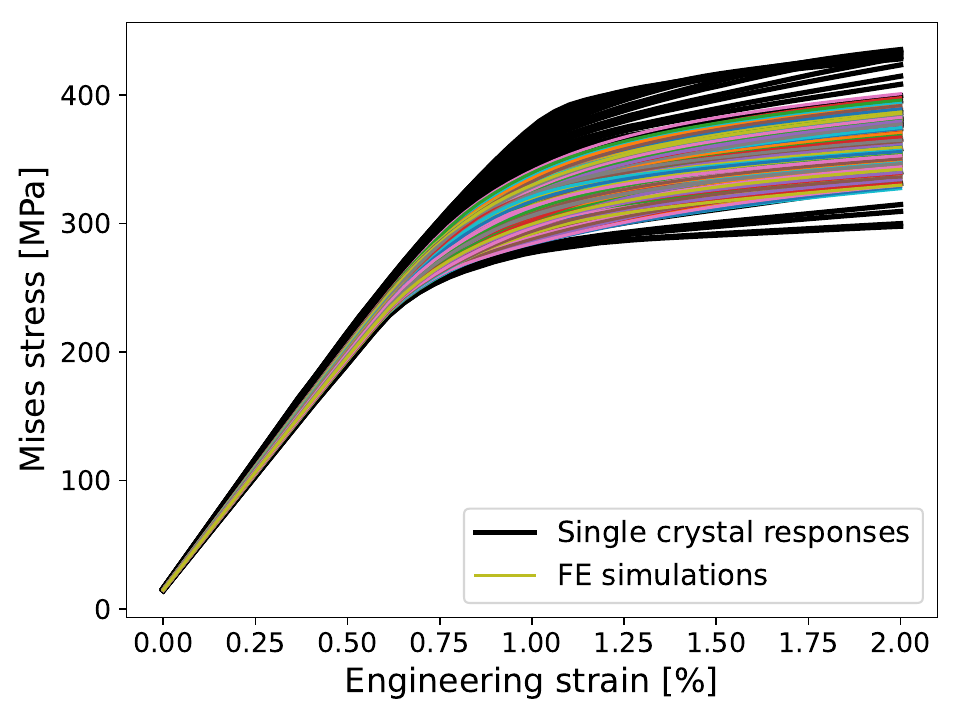}
         \label{c2}
     }
     \subfloat[]{
         \includegraphics[trim={0cm 0cm 0cm 0cm},clip,width=0.245\textwidth]{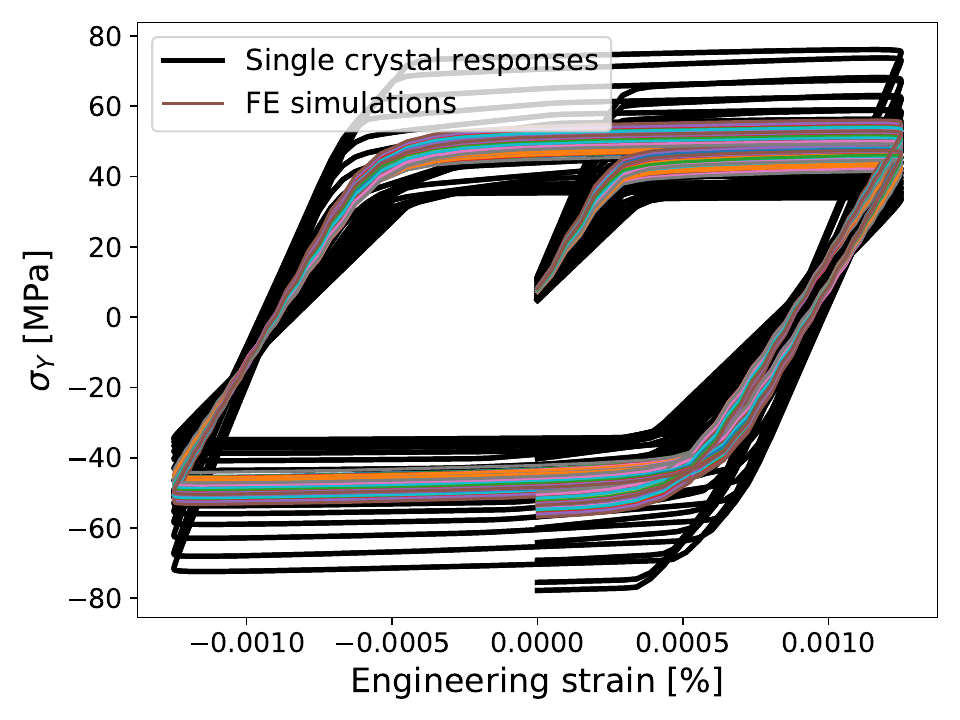}
         \label{c4}
     }
    \caption{\psubref{c1} Aluminum under 1\% tension. \psubref{c3} Copper under 0.125\% tension. \psubref{c2} Aluminum under 2\% shear. \psubref{c4} Copper under cyclic loading. }
    \label{response_curves}
\end{figure}

The idea of using single crystal responses as "basis functions" to obtain the RVE responses via certain nonlinear rule of mixture relies on the assumption that the different RVE responses are bounded within the variation of the single crystal responses given a similar strain state. To check the validity of this assumption, we plotted the single crystal responses against all the polycrystal RVE responses for the respective cases in \fref{response_curves}. 

From the curves shown in \fref{response_curves}, it is evident that there are significant variations in polycrystal RVE responses for each case due to local microstructure changes. It is also observed that the stress-strain curves from different cases drastically differ in shape and magnitude due to changing material properties and/or loading conditions. However, despite the significant variations, all polycrystal RVE responses fall entirely within the variations of the single crystal responses (i.e., black lines in \fref{response_curves}), therefore showing the validity of our assumption. It is also emphasized that the single crystals and the RVEs should be subject to identical displacement BCs that lead to a similar mean strain state, since strain (as a function of time) was not an direct input to the DeepONet, and only the stress values at different time steps were fed into the DeepONet. Hence, it is required that the data points at the same time step share an similar strain state.

\subsection{DeepONet performance for a single material and loading}
\label{sec:baseline}
Before conducting transfer learning, both DeepONets were first trained on a set of aluminum data, which was under 1\% tensile strain. 1000 simulations were conducted, and 80\% of the data was used in training while 20\% was reserved for testing. To test the repeatability of the results, each model was independently trained 4 times with randomly generated training/testing data split. Out of the 4 repetitions, the baseline MP-DeepONet achieved an averaged relative error of 3.02\%, while that of the proposed SC-DeepONet was 1.45\%. Training times for both networks were similar, with an average training time of 429s and 445s, respectively. The inference time per case was similar for both models, at around $8\times10^{-3}$s. On average, running each CP simulation took 196.7s, making the inference 24586 times faster than the direct numerical simulation. Detailed performance metrics for the best-performing run out of the four are shown in \tref{metrics_base}.
\begin{table}[h!]
    \caption{Performance metrics for the aluminum under 1\% tension}
    \centering
    \begin{tabular}{ccccc}
     Model & \vline & Relative error [\%] & Mean absolute error [MPa] & $R^2$ value\\
    \hline
    MP-DeepONet & \vline  & 2.61 & 9.97 & 0.978 \\
    SC-DeepONet & \vline  & 1.06 & 3.62 & 0.997 \\
    \end{tabular}
    \label{metrics_base}
\end{table}

\begin{figure}[h!] 
    \centering
     \subfloat[]{
         \includegraphics[trim={0cm 0cm 0cm 0cm},clip,width=0.31\textwidth]{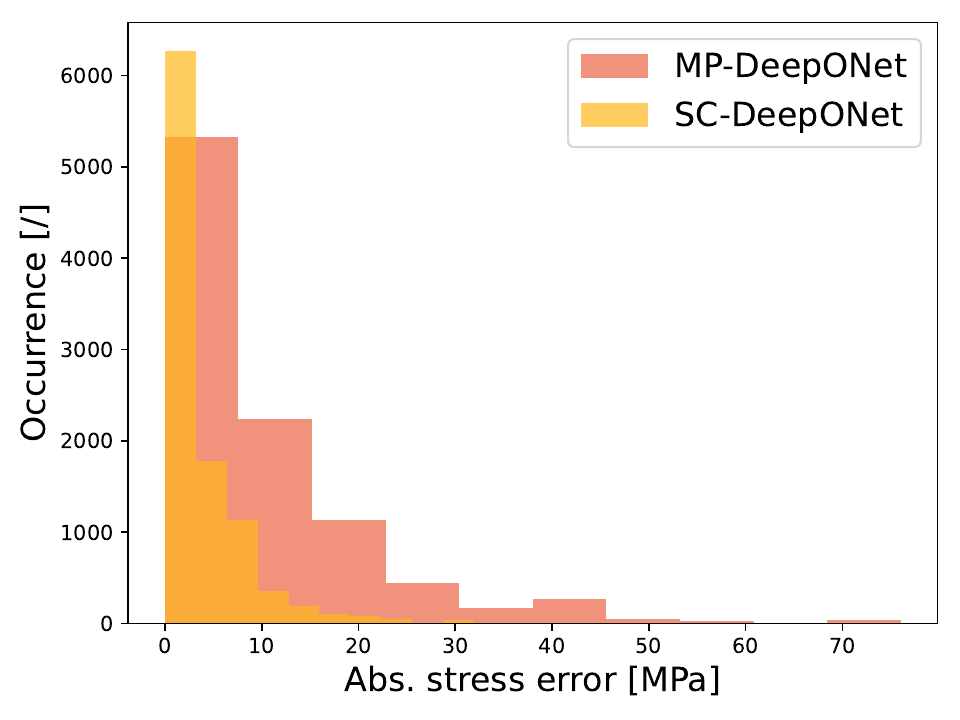}
         \label{hist}
     }
     \subfloat[]{
         \includegraphics[trim={0cm 0cm 0cm 0cm},clip,width=0.31\textwidth]{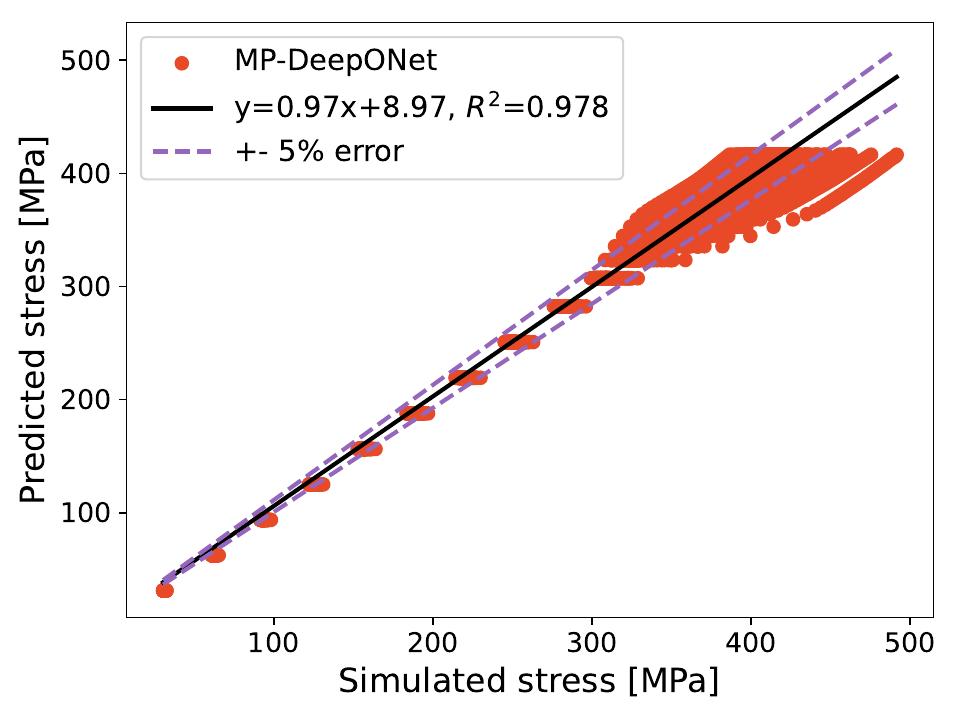}
         \label{mp_corr}
     }
     \subfloat[]{
         \includegraphics[trim={0cm 0cm 0cm 0cm},clip,width=0.31\textwidth]{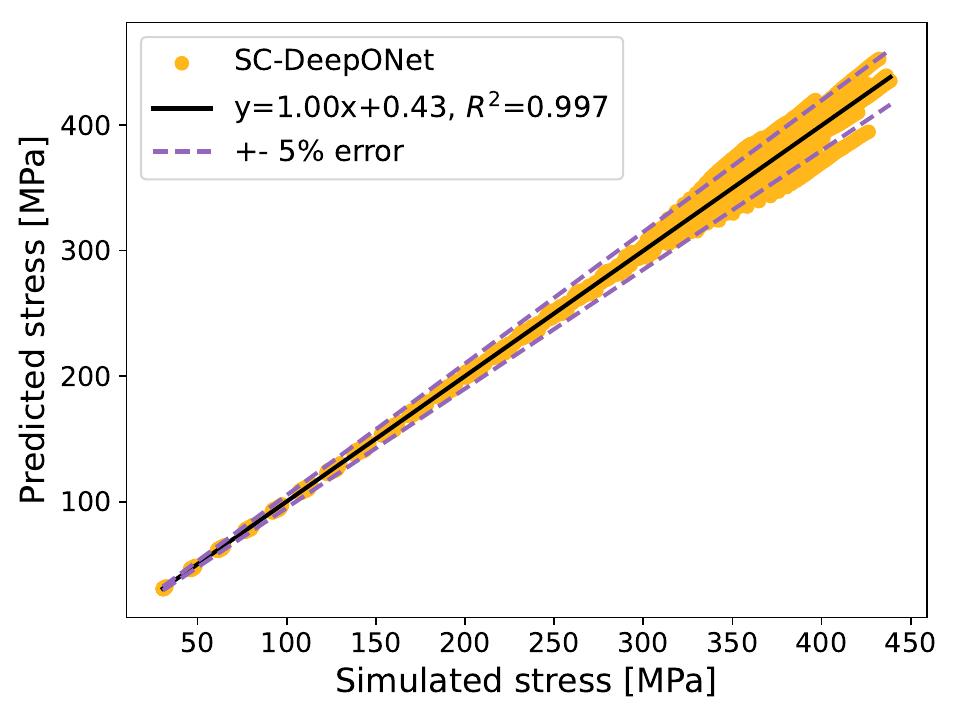}
         \label{sc_corr}
     }
    \caption{\psubref{hist} Histogram of absolute stress error. Scatter plot comparing simulated and predicted stresses for: \psubref{mp_corr} MP-DeepONet, \psubref{sc_corr} SC-DeepONet. }
    \label{al_tension_1}
\end{figure}

\begin{figure}[h!]
\newcommand\x{0.23}
    \centering
    \begin{tabular}{ c c c c c }
    \begin{minipage}[c]{\x\textwidth}
       \centering 
        \subfloat[MP-DeepONet, best]{\includegraphics[trim={0cm 0cm 0cm 0cm},clip,width=\textwidth]{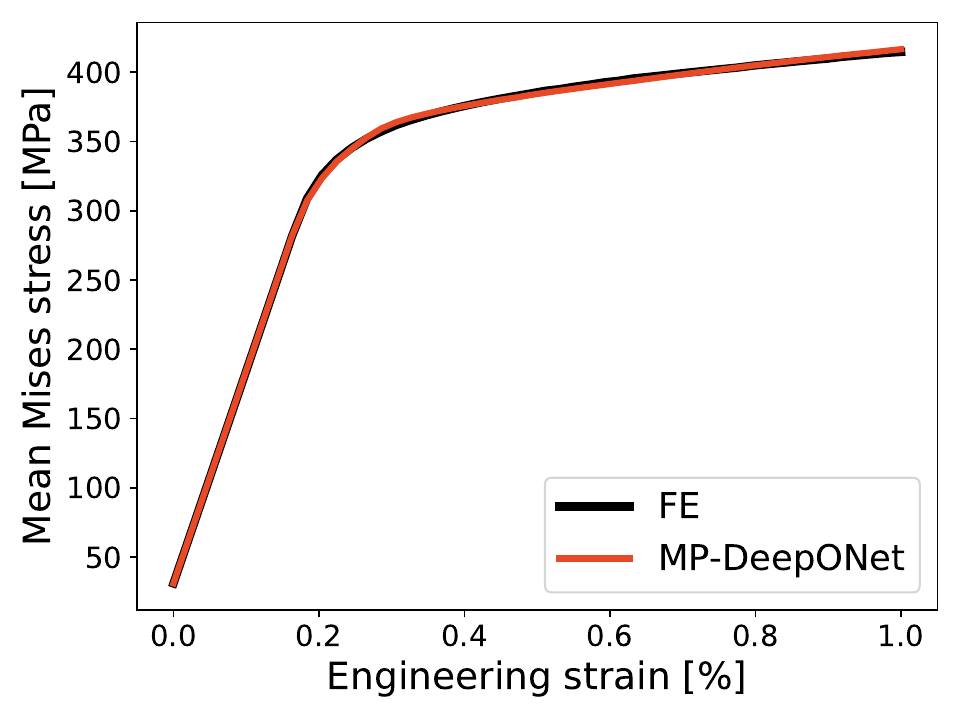}
        \label{p11}}
    \end{minipage} &
    \begin{minipage}[c]{\x\textwidth}
       \centering 
        \subfloat[MP-DeepONet, 50$^{th}$ pct]{\includegraphics[trim={0cm 0cm 0cm 0cm},clip,width=\textwidth]{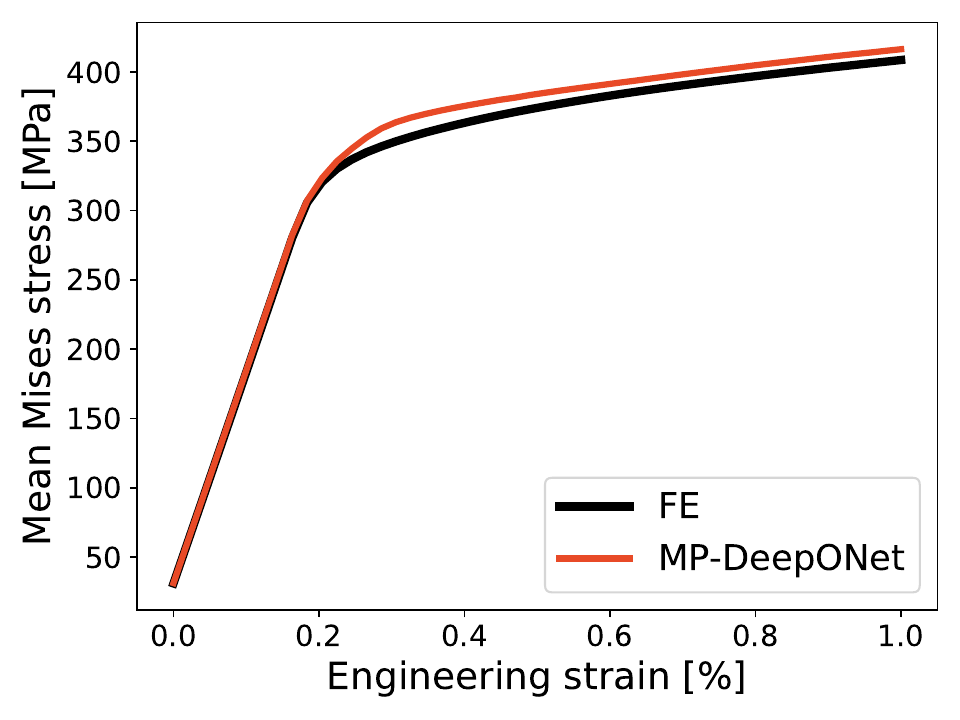}
        \label{p12}}
    \end{minipage} &
    \begin{minipage}[c]{\x\textwidth}
       \centering 
        \subfloat[MP-DeepONet, 85$^{th}$ pct]{\includegraphics[trim={0cm 0cm 0cm 0cm},clip,width=\textwidth]{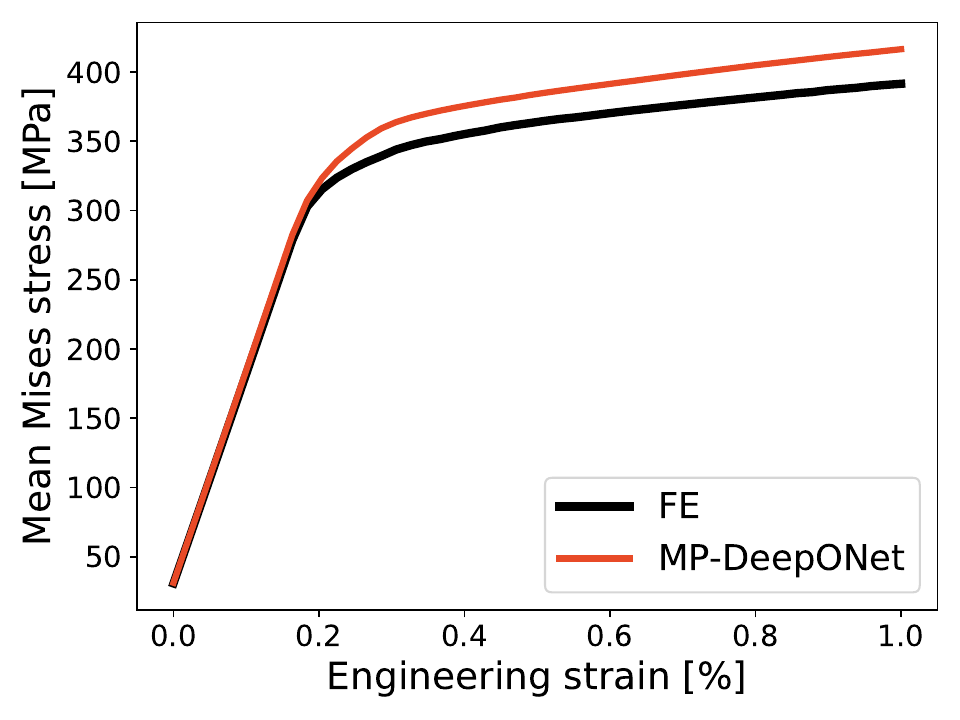}
        \label{p13}}
    \end{minipage} &
    \begin{minipage}[c]{\x\textwidth}
       \centering 
        \subfloat[MP-DeepONet, worst]{\includegraphics[trim={0cm 0cm 0cm 0cm},clip,width=\textwidth]{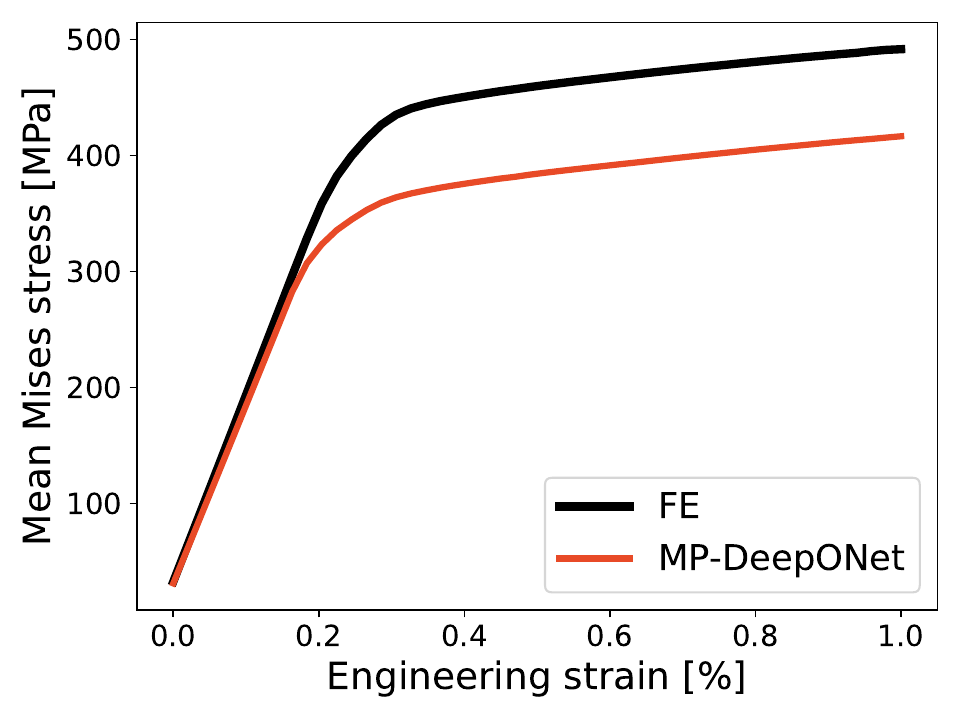}
        \label{p14}}
    \end{minipage} \\

    \begin{minipage}[c]{\x\textwidth}
       \centering 
        \subfloat[SC-DeepONet, best]{\includegraphics[trim={0cm 0cm 0cm 0cm},clip,width=\textwidth]{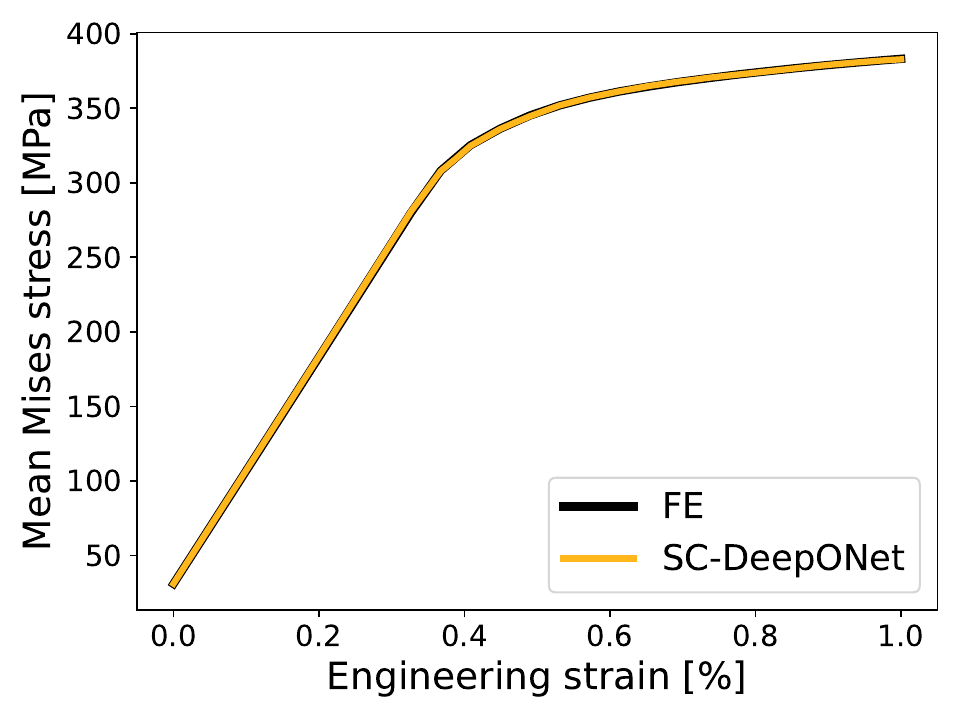}
        \label{p21}}
    \end{minipage} &
    \begin{minipage}[c]{\x\textwidth}
       \centering 
        \subfloat[SC-DeepONet, 50$^{th}$ pct]{\includegraphics[trim={0cm 0cm 0cm 0cm},clip,width=\textwidth]{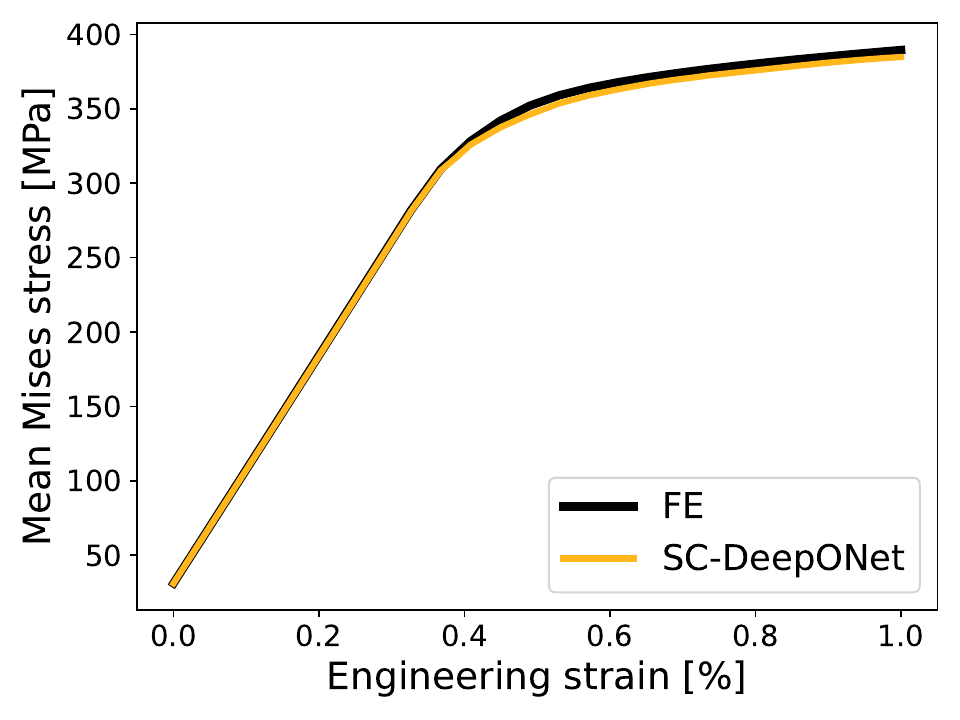}
        \label{p22}}
    \end{minipage} &
    \begin{minipage}[c]{\x\textwidth}
       \centering 
        \subfloat[SC-DeepONet, 85$^{th}$ pct]{\includegraphics[trim={0cm 0cm 0cm 0cm},clip,width=\textwidth]{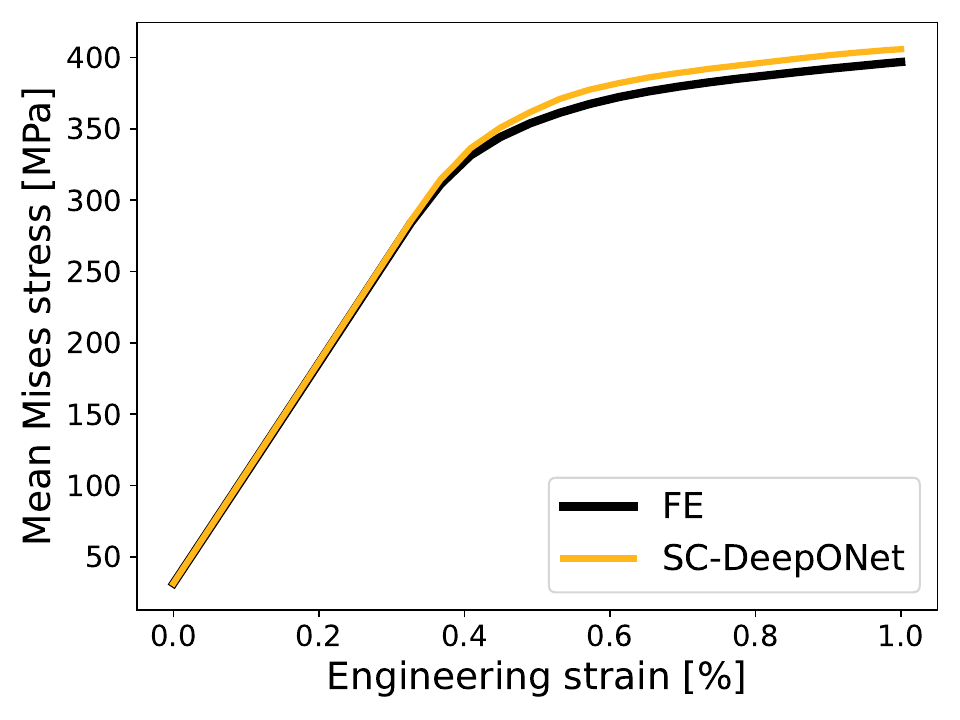}
        \label{p23}}
    \end{minipage} &
    \begin{minipage}[c]{\x\textwidth}
       \centering 
        \subfloat[SC-DeepONet, worst]{\includegraphics[trim={0cm 0cm 0cm 0cm},clip,width=\textwidth]{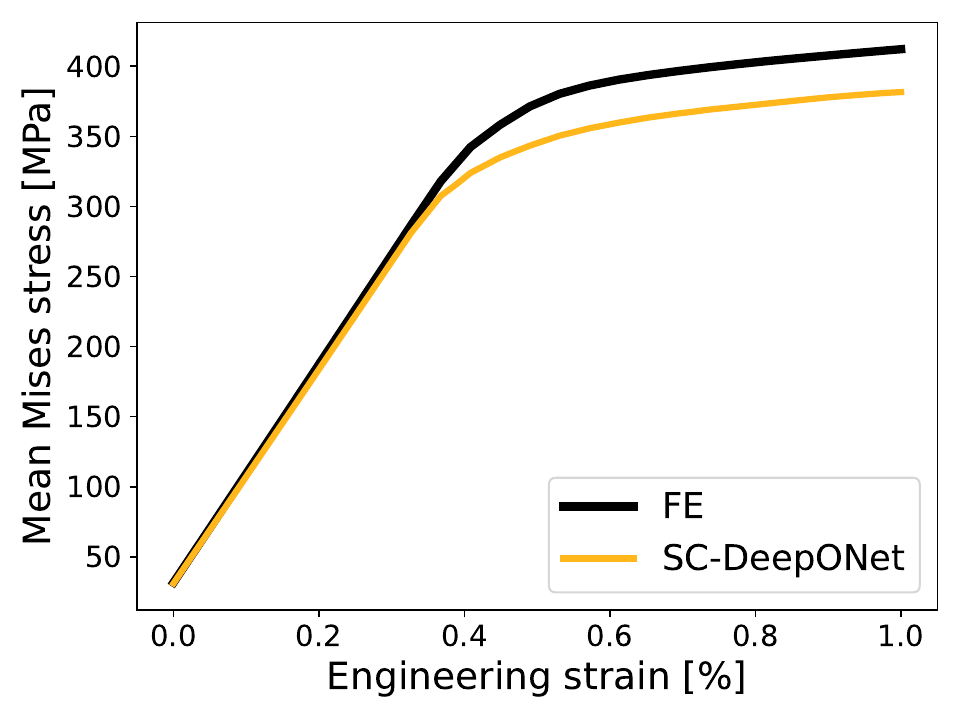}
        \label{p24}}
    \end{minipage} \\

    \end{tabular}
    \caption{Testing results for the two models at different error percentiles, for aluminum under 1\% tension.}
    \label{al_tension}
\end{figure}

To further compare the distribution of prediction errors of the MP- and SC-DeepONets, histogram and scatter plots are shown in \fref{al_tension_1}. Selected stress-strain curves are shown in \fref{al_tension}. They are ranked by the percentile of mean relative error in the predictions, and the 0$^{th}$ (best case),  50$^{th}$ (median case), 85$^{th}$ and 100$^{th}$ (worst case) percentiles are shown for statistical representation. 

From the results, we observe that the SC-DeepONet could provide a higher prediction accuracy while keeping a training time similar to that of the MP-DeepONet. Both networks could efficiently predict the mean stress-strain curve at a rate 5 orders of magnitude faster than FE simulations. It is also worth mentioning that the 36 single crystal responses were obtained in merely 5s (total time), which was only 2.5\% the time of one RVE simulation. Note that the single crystal responses are only generated \emph{once} for this material and loading and do not change with the RVE microstructure and the number of training data points. Therefore, the computational expense associated with using the SC-DeepONet is negligible. For MP-DeepONet, 85.6\% of the time step predictions have a relative error $\le 5\%$, while that for SC-DeepONet is 98.1\%. The stress-strain plots in \fref{al_tension} show that, in the best case, both networks can faithfully predict the mean stress-strain curve of the RVE. However, SC-DeepONet continues to deliver accurate predictions with minimal deviation up to that 85$^{th}$ percentile case, while the MP-DeepONet already shows significant over-prediction. In the worst case, both networks accurately predicted the elastic response and the hardening slope but under-predicted the initial yield stress.

\subsection{Generalization to different materials and strains}
\label{sec:cu}
\begin{figure}[h!] 
    \centering
     \subfloat[]{
         \includegraphics[trim={0cm 0cm 0cm 0cm},clip,width=0.31\textwidth]{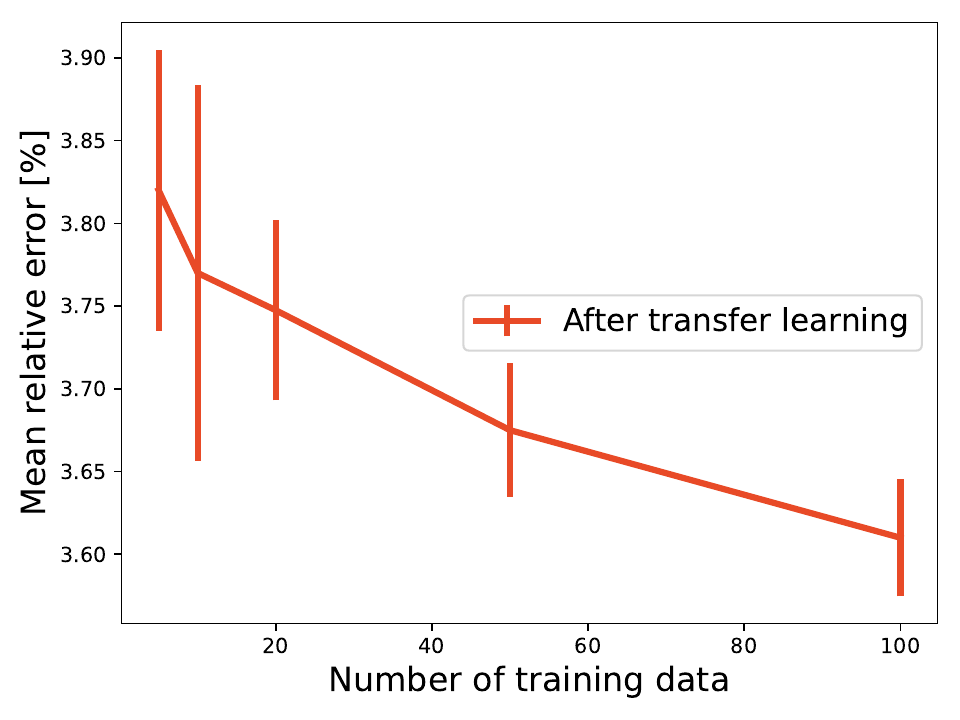}
         \label{tr_mp_cu}
     }
     \subfloat[]{
         \includegraphics[trim={0cm 0cm 0cm 0cm},clip,width=0.31\textwidth]{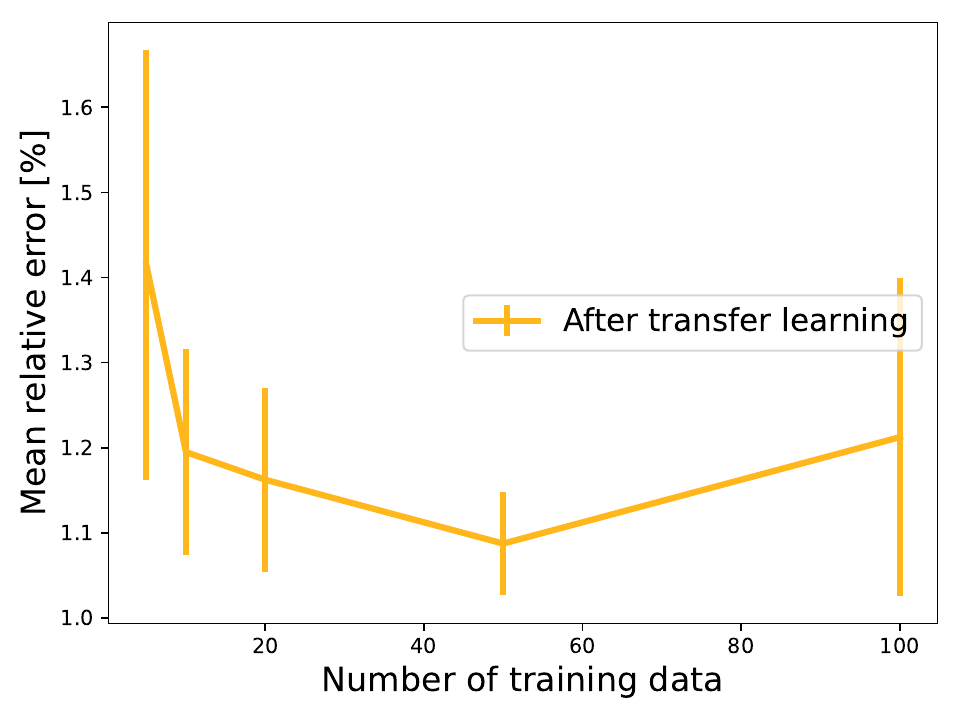}
         \label{tr_sc_cu}
     }
     \subfloat[]{
         \includegraphics[trim={0cm 0cm 0cm 0cm},clip,width=0.31\textwidth]{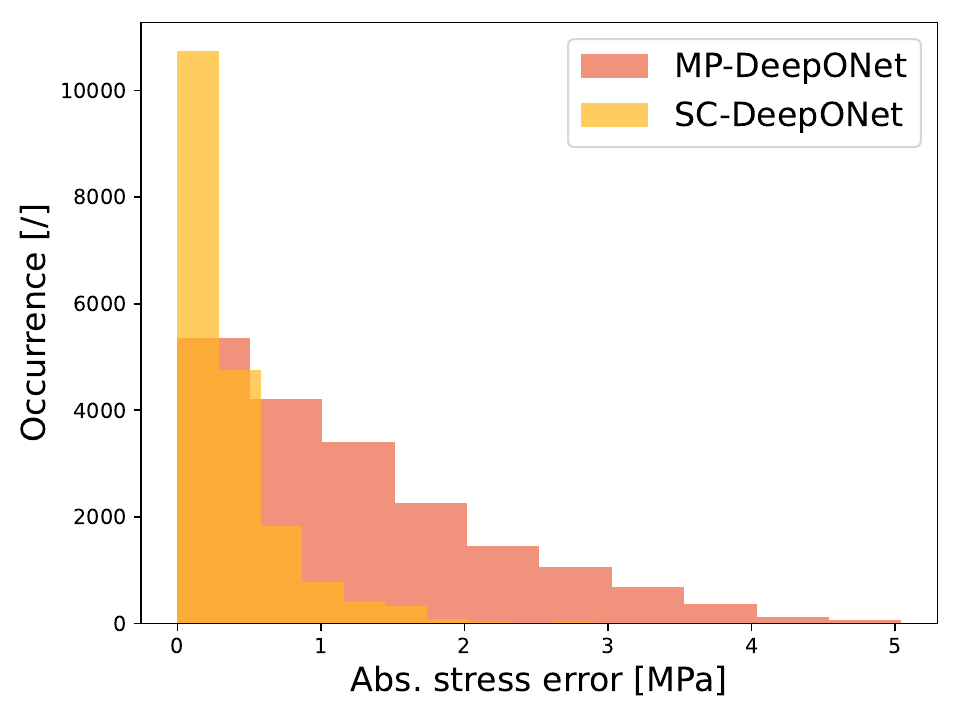}
         \label{tr_cu_hist}
     }
    \caption{Final prediction error for different number of transfer learning data points: \psubref{tr_mp_cu} MP-DeepONet and \psubref{tr_sc_cu} SC-DeepONet. \psubref{tr_cu_hist} Error histogram for both DeepONets after transfer learning with 20 data points.}
    \label{cu_1}
\end{figure}

\begin{figure}[h!]
\newcommand\x{0.23}
    \centering
    \begin{tabular}{ c c c c c }
    \begin{minipage}[c]{\x\textwidth}
       \centering 
        \subfloat[MP-DeepONet, best]{\includegraphics[trim={0cm 0cm 0cm 0cm},clip,width=\textwidth]{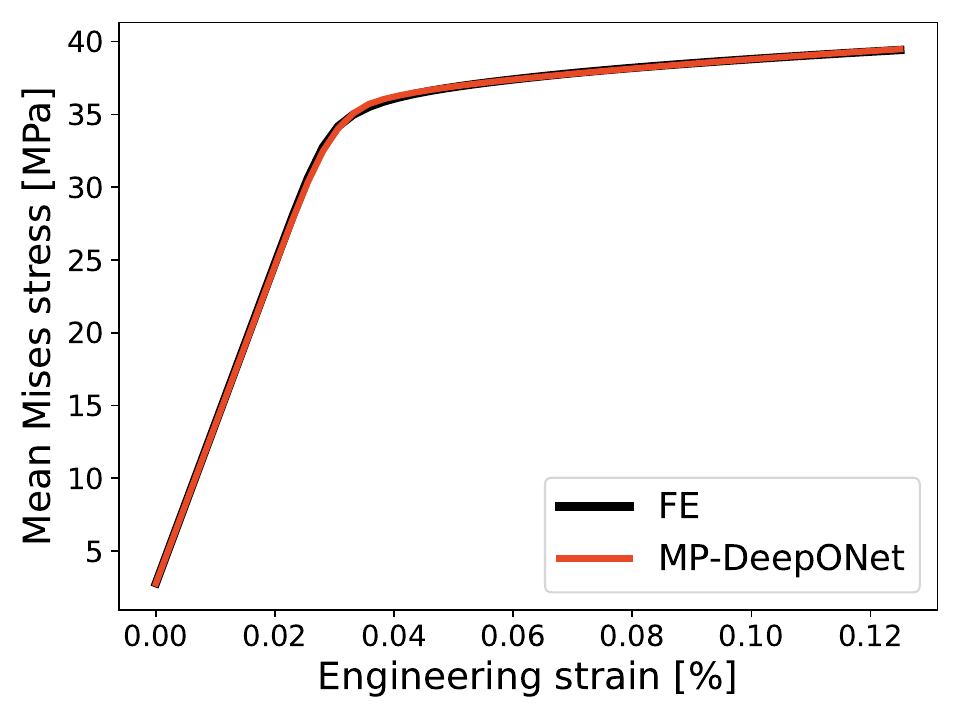}
        \label{p31}}
    \end{minipage} &
    \begin{minipage}[c]{\x\textwidth}
       \centering 
        \subfloat[MP-DeepONet, 50$^{th}$ pct]{\includegraphics[trim={0cm 0cm 0cm 0cm},clip,width=\textwidth]{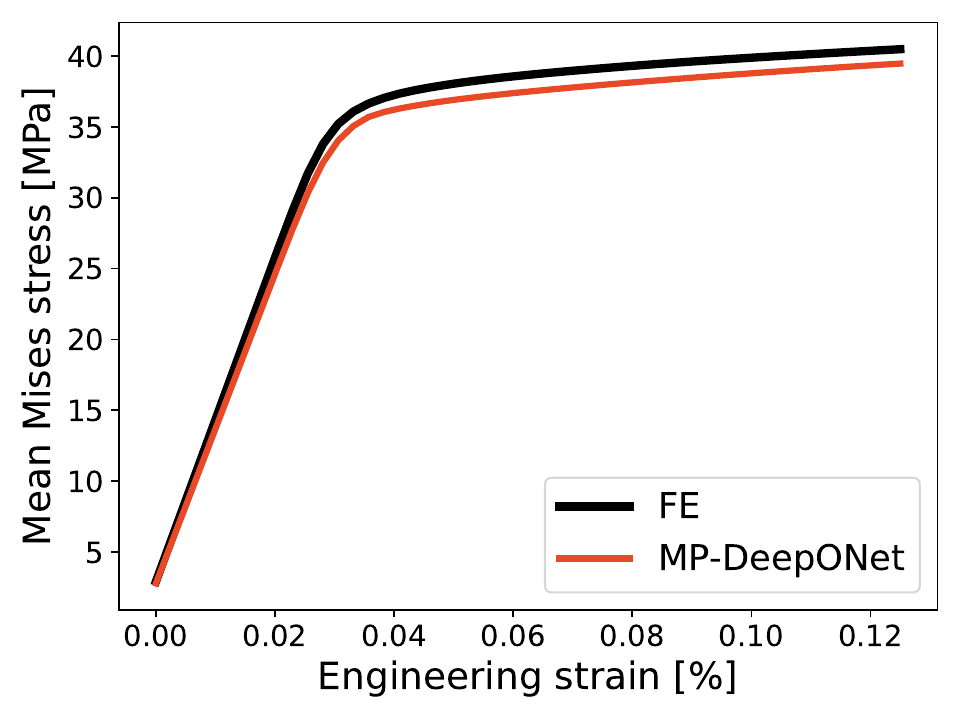}
        \label{p32}}
    \end{minipage} &
    \begin{minipage}[c]{\x\textwidth}
       \centering 
        \subfloat[MP-DeepONet, 85$^{th}$ pct]{\includegraphics[trim={0cm 0cm 0cm 0cm},clip,width=\textwidth]{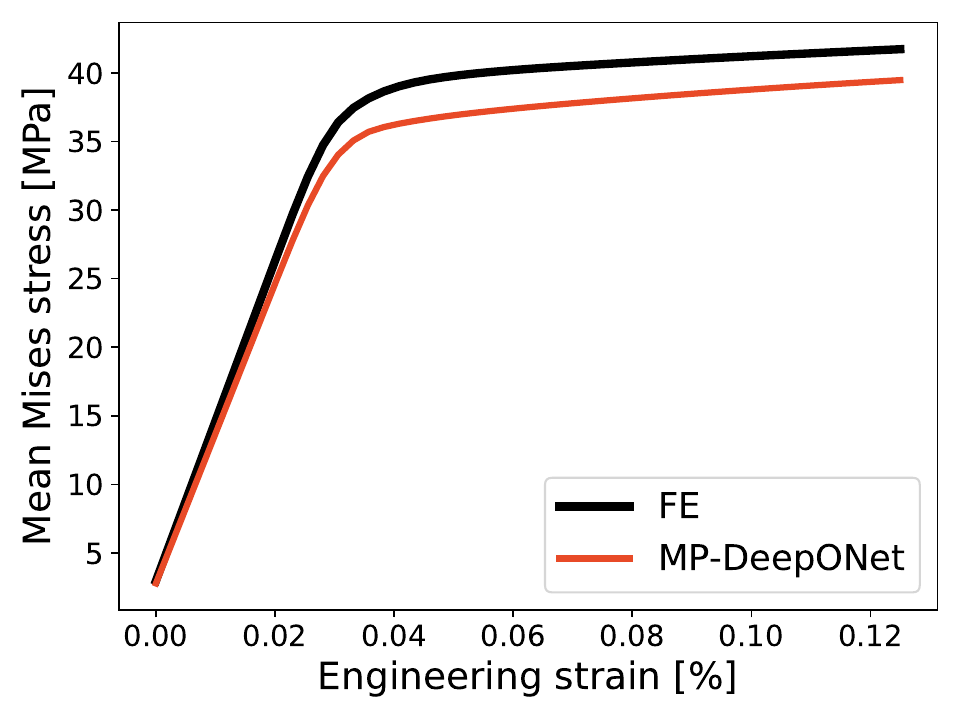}
        \label{p33}}
    \end{minipage} &
    \begin{minipage}[c]{\x\textwidth}
       \centering 
        \subfloat[MP-DeepONet, worst]{\includegraphics[trim={0cm 0cm 0cm 0cm},clip,width=\textwidth]{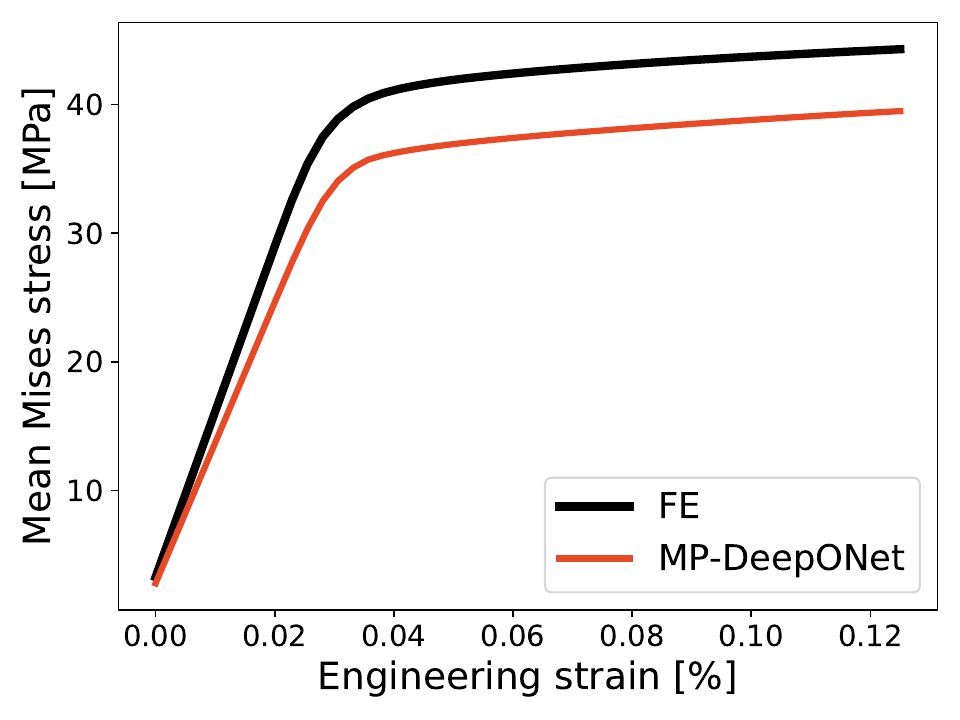}
        \label{p34}}
    \end{minipage} \\

    \begin{minipage}[c]{\x\textwidth}
       \centering 
        \subfloat[SC-DeepONet, best]{\includegraphics[trim={0cm 0cm 0cm 0cm},clip,width=\textwidth]{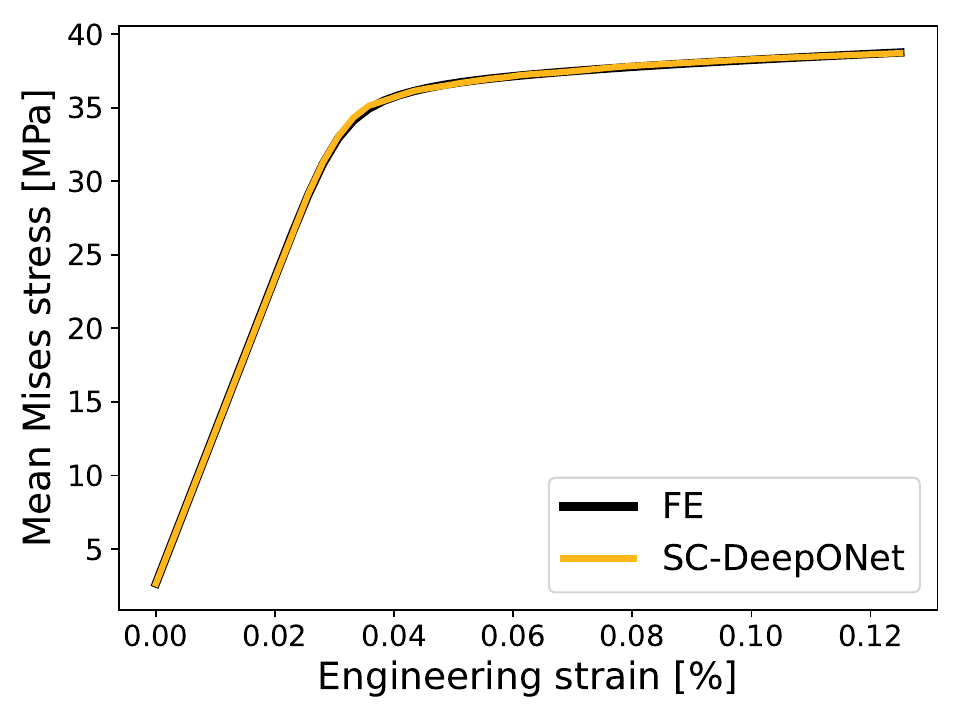}
        \label{p41}}
    \end{minipage} &
    \begin{minipage}[c]{\x\textwidth}
       \centering 
        \subfloat[SC-DeepONet, 50$^{th}$ pct]{\includegraphics[trim={0cm 0cm 0cm 0cm},clip,width=\textwidth]{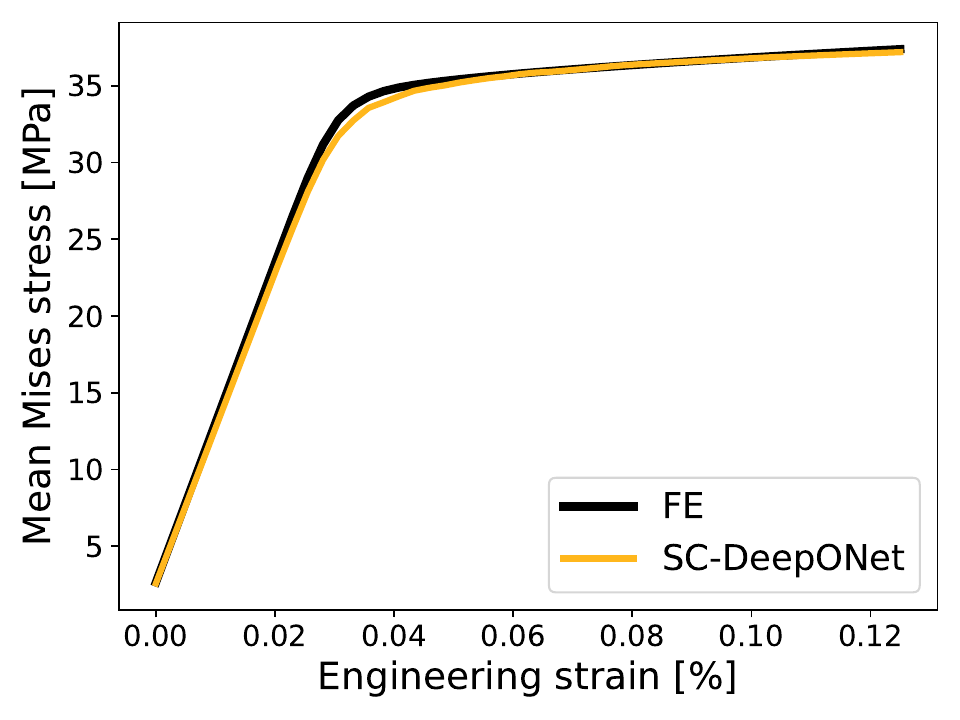}
        \label{p42}}
    \end{minipage} &
    \begin{minipage}[c]{\x\textwidth}
       \centering 
        \subfloat[SC-DeepONet, 85$^{th}$ pct]{\includegraphics[trim={0cm 0cm 0cm 0cm},clip,width=\textwidth]{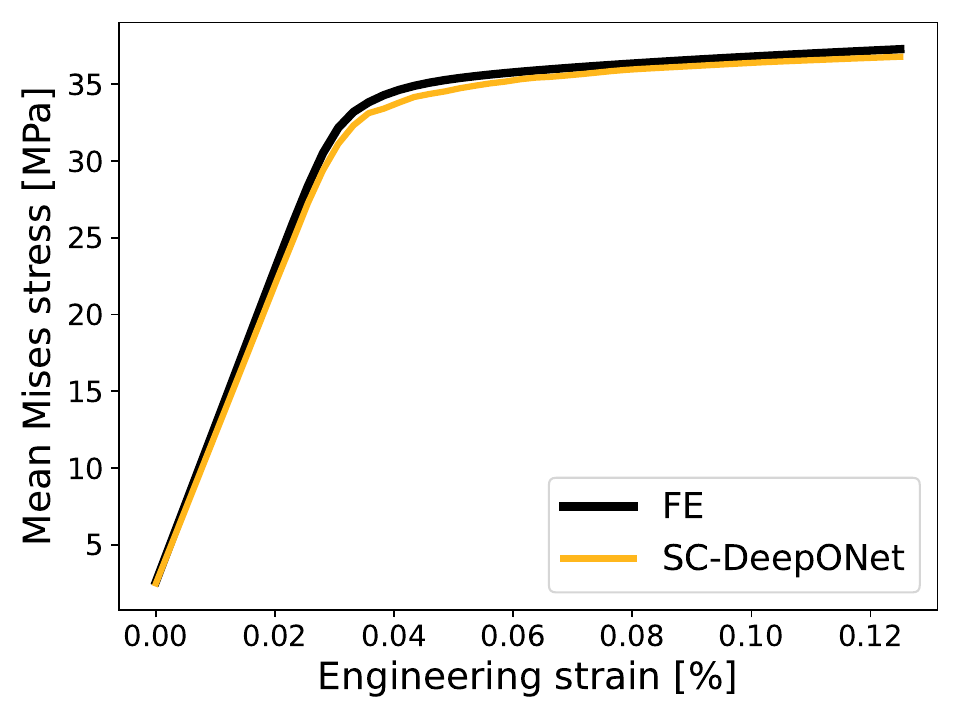}
        \label{p43}}
    \end{minipage} &
    \begin{minipage}[c]{\x\textwidth}
       \centering 
        \subfloat[SC-DeepONet, worst]{\includegraphics[trim={0cm 0cm 0cm 0cm},clip,width=\textwidth]{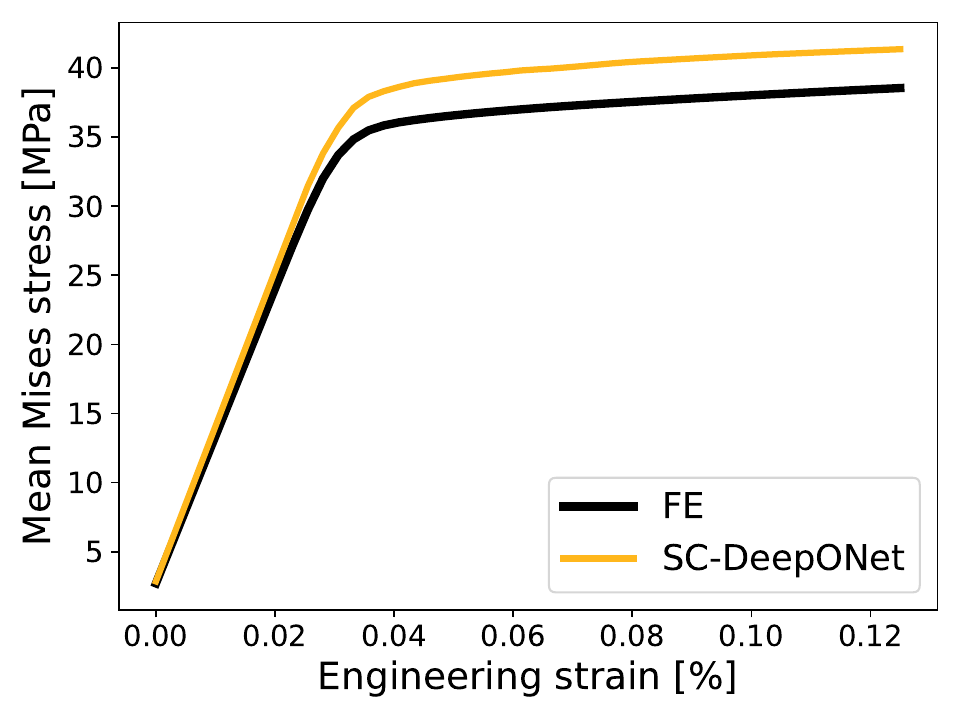}
        \label{p44}}
    \end{minipage} \\

    \end{tabular}
    \caption{Testing results for the two models at different error percentiles after transfer learning with 20 data points, for copper under 0.125\% tension.}
    \label{cu_tension}
\end{figure}

In \sref{sec:baseline}, we demonstrated that different DeepONets could be conditioned to a single material property/loading state and provide reasonable predictions. We now investigate how MP- and SC-DeepONets can be extended to different materials and strain levels. In this section, we considered the case of copper under 0.125\% tension, for which case 400 simulation results were generated. To investigate the effect of the number of copper data points on the transfer learning accuracy, five different numbers of data were used in the fine-tuning: [5, 10, 20, 50, 100]. For repeatability, each training was repeated 4 times. Training times with 20 new data points were similar for MP- and SC-DeepONets and were close to 45s. Once trained, the prediction time for both models remained similar to the baseline case. The average CP simulation time was 75.5s, making the DeepONet predictions 9436 times faster. Generating the single crystal responses took merely 4s, which is quick compared to the time of a single RVE simulation. The transfer learning results with different data points are shown in \fref{tr_mp_cu} and \fref{tr_sc_cu}. The models fine-tuned using 20 new data points were chosen for further inspection, and the histogram of the prediction errors is shown in \fref{tr_cu_hist}. Key accuracy metrics are presented in \tref{metrics_cu}. The predicted stress-strain curves from the two models at different error percentiles are depicted in \fref{cu_tension}.
\begin{table}[h!]
    \caption{Performance metrics for the copper under 0.125\% tension}
    \centering
    \begin{tabular}{cccccc}
     Model & \vline & Rel. err. (before) [\%] & Rel. err. (after) [\%] & Mean abs. err. [MPa] & $R^2$ \\
    \hline
    MP-DeepONet & \vline  & 2.92$\times10^6$ & 3.82 & 1.22 & 0.975 \\
    SC-DeepONet & \vline  & 2.87 & 1.11 & 0.36 & 0.997 \\
    \end{tabular}
    \label{metrics_cu}
\end{table}

The transfer learning results in \tref{metrics_cu} reveal more differences in the MP- and SC-DeepONets. We see that for MP-DeepONet, the prediction error on copper data before any transfer learning is high, indicating that the information the network learned from the previous aluminum data is not directly generalizable. However, in the case of SC-DeepONet, with the help of the new single crystal basis functions for copper under tension, the prediction error is already less than 3\% before any transfer learning, indicating that the combination of local microstructure information (extracted from the ResUNet trunk network) and material response information (provided to the branch as new single crystal responses) is effective and highly generalizable. For both networks, prediction performance generally increased with more copper data added to the training. However, the SC-DeepONet provided higher accuracy than MP-DeepONet for all numbers of copper training data. With merely 20 copper RVE data used in fine-tuning, the SC-DeepONet achieved close to 1\% relative error, and the mean absolute error in stress was less than 0.5 MPa, showing extreme data efficiency and generalizability of the proposed architecture. When inspecting the predicted stress-strain curves in \fref{cu_tension}, we see that the SC-DeepONet predicted behavior matched closely with the CP simulations up to 85$^{th}$ percentile. Another vital advantage of the SC-DeepONet can be seen by comparing the worst-case predictions of the two networks. Comparing the overall RVE response for aluminum (\fref{c1}) and copper (\fref{c2}), it is immediately apparent that the copper material displays more elastic anisotropy. From the last column of \fref{cu_tension}, we see that the SC-DeepONet, with the help of the anisotropic single crystal responses, can accurately capture the elastic slope even in the worst case. At the same time, the MP-DeepONet struggles to capture the initial modulus when the elastic responses vary significantly in the test set due to material anisotropy. These observations prove that using single crystal stress-strain curves as inputs to the branch network is a more mechanically meaningful way to capture material behavior than directly feeding the network with material property values.

\subsection{Generalization to different boundary conditions}
\label{sec:shear}
\begin{figure}[h!] 
    \centering
     \subfloat[]{
         \includegraphics[trim={0cm 0cm 0cm 0cm},clip,width=0.31\textwidth]{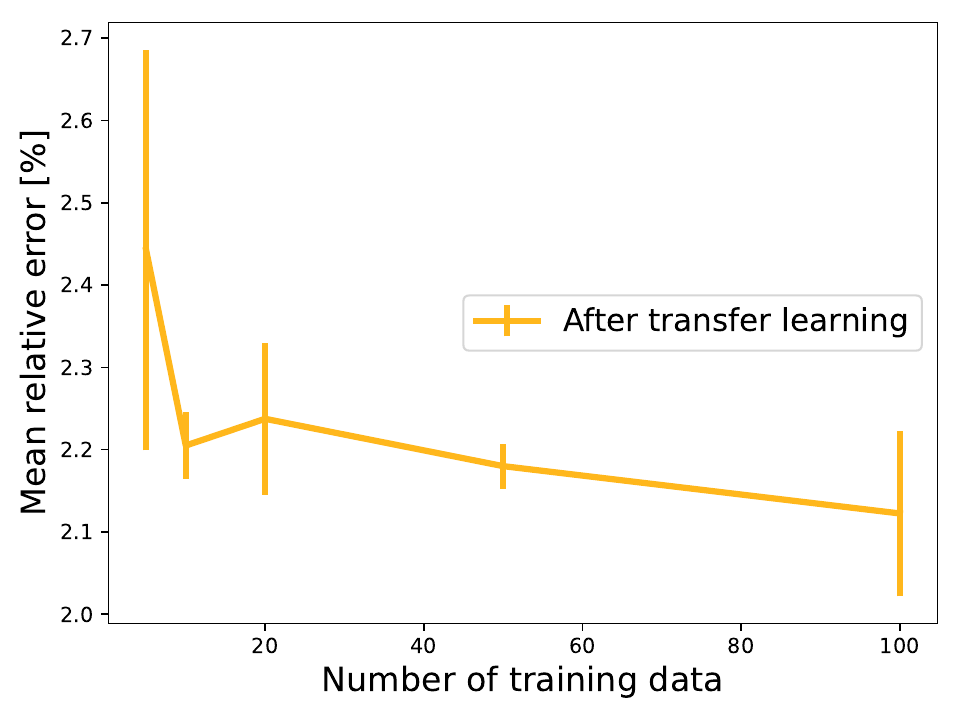}
         \label{tr_sh}
     }
     \subfloat[]{
         \includegraphics[trim={0cm 0cm 0cm 0cm},clip,width=0.31\textwidth]{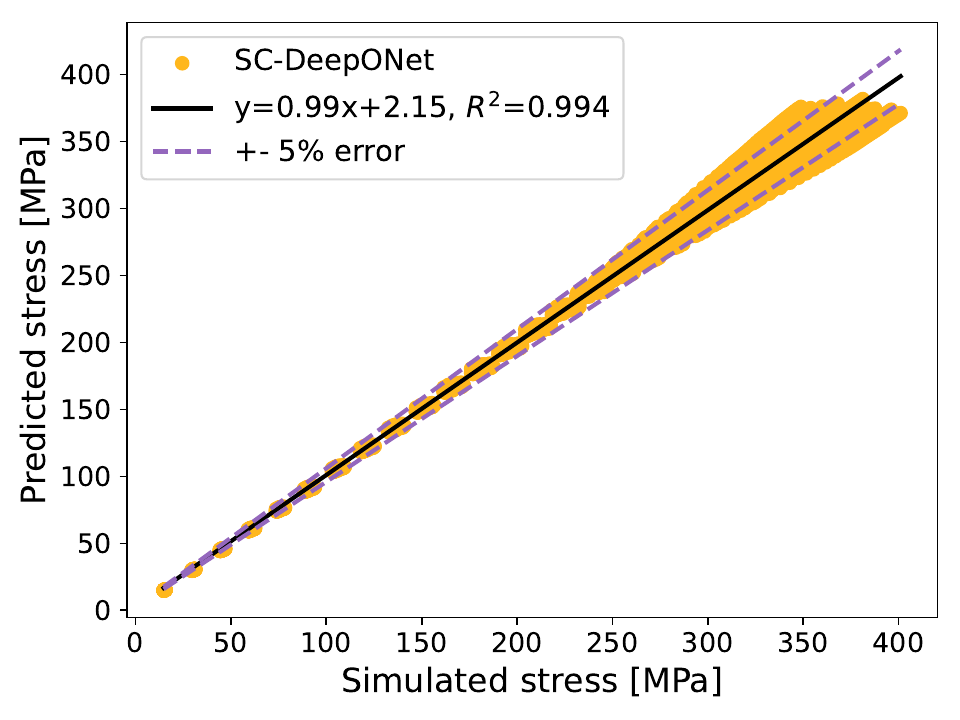}
         \label{tr_sh_corr}
     }
    \caption{\psubref{tr_sh} Final SC-DeepONet prediction error for different numbers of transfer learning data points. \psubref{tr_sh_corr} Correlation of predicted and FE simulated stress data.}
    \label{al_shear1}
\end{figure}

\begin{figure}[h!]
\newcommand\x{0.23}
    \centering
    \begin{tabular}{ c c c c c }
    \begin{minipage}[c]{\x\textwidth}
       \centering 
        \subfloat[SC-DeepONet, best]{\includegraphics[trim={0cm 0cm 0cm 0cm},clip,width=\textwidth]{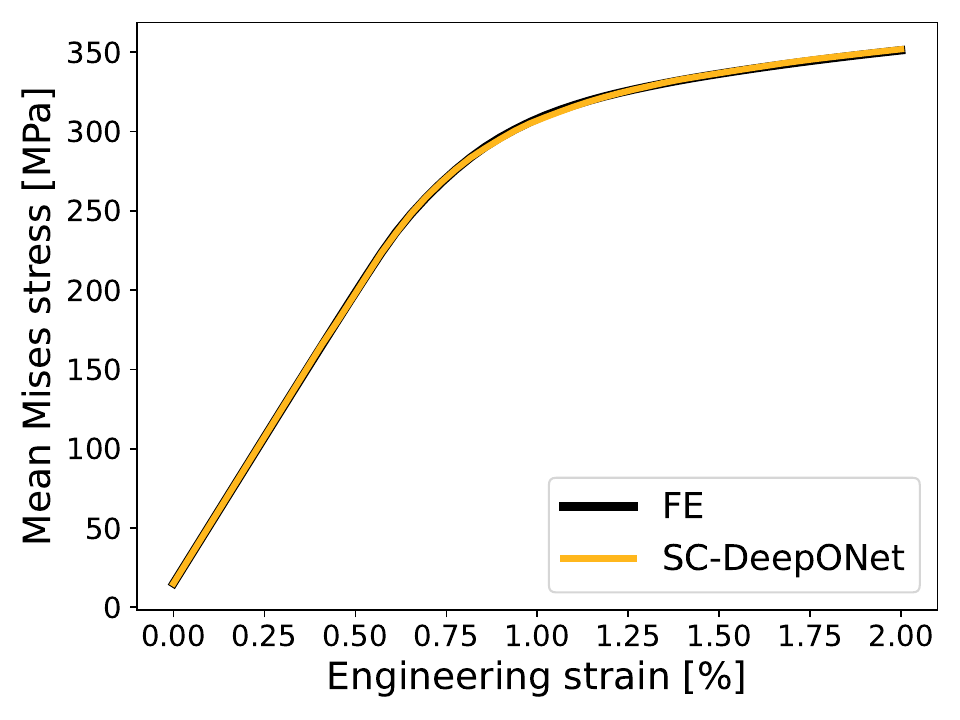}
        \label{p51}}
    \end{minipage} &
    \begin{minipage}[c]{\x\textwidth}
       \centering 
        \subfloat[SC-DeepONet, 50$^{th}$ pct]{\includegraphics[trim={0cm 0cm 0cm 0cm},clip,width=\textwidth]{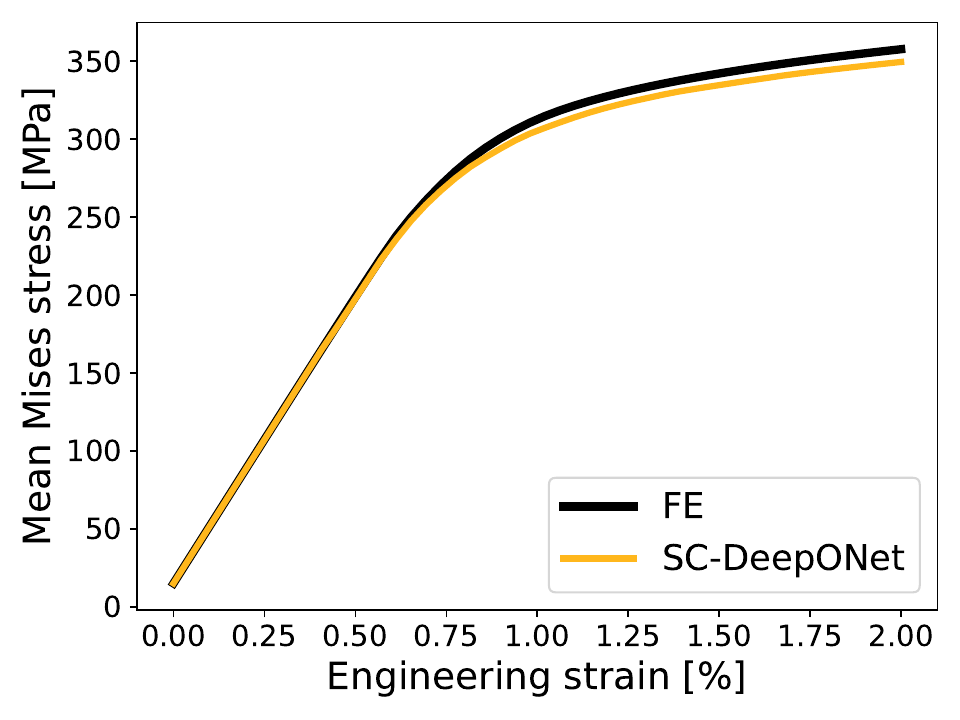}
        \label{p52}}
    \end{minipage} &
    \begin{minipage}[c]{\x\textwidth}
       \centering 
        \subfloat[SC-DeepONet, 85$^{th}$ pct]{\includegraphics[trim={0cm 0cm 0cm 0cm},clip,width=\textwidth]{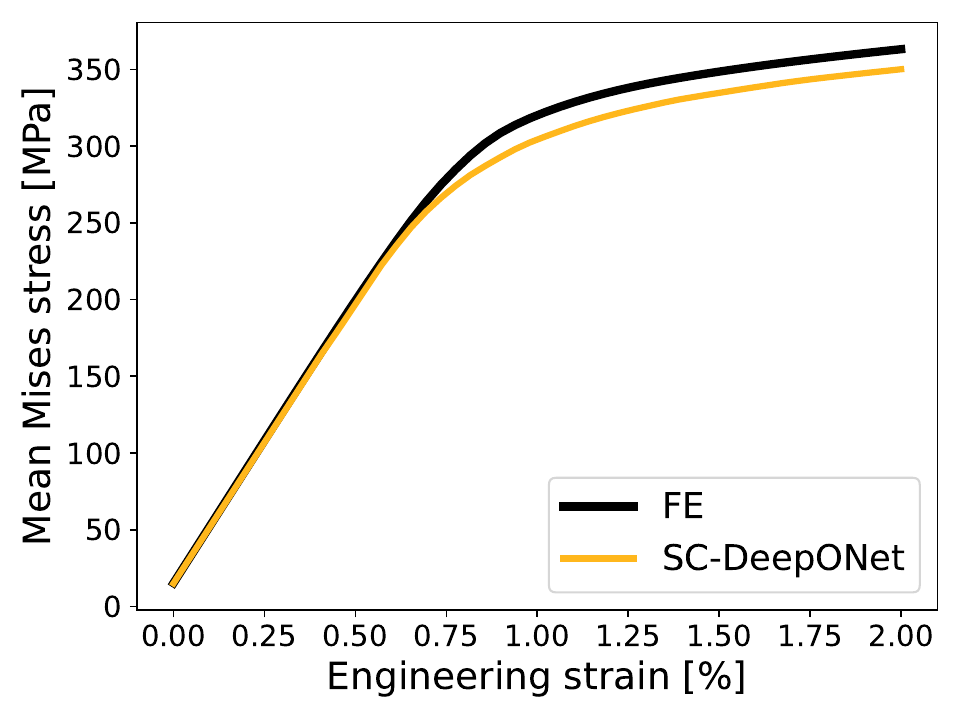}
        \label{p53}}
    \end{minipage} &
    \begin{minipage}[c]{\x\textwidth}
       \centering 
        \subfloat[SC-DeepONet, worst]{\includegraphics[trim={0cm 0cm 0cm 0cm},clip,width=\textwidth]{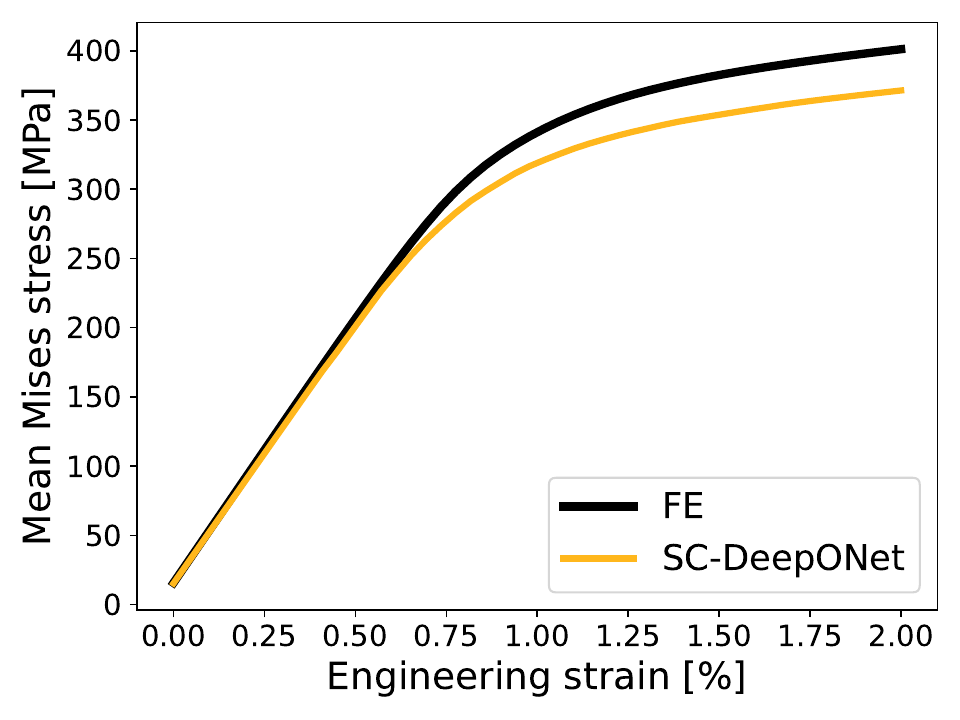}
        \label{p54}}
    \end{minipage} \\

    \end{tabular}
    \caption{Testing results for SC-DeepONet at different error percentiles after transfer learning with 20 data points, for aluminum under 2\% shear.}
    \label{al_shear}
\end{figure}

In this section, we move away from tensile loading and investigate SC-DeepONet generalizability for aluminum under shear loading. 400 RVE simulations were conducted, and new single-crystal responses were generated. Contrary to the tensile cases, where those responses were generated from a single finite element, for shear, a uniform mesh of 10$\times$10 elements was used to model the single crystals and obtained their responses \footnote{If the simple shear boundary conditions are applied on a single finite element, it will lead to a state of uniform shear strain since all the degrees of freedom are prescribed. This is not the case for an inhomogeneous polycrystal consisting of multiple elements since it has additional unconstrained degrees of freedom that can deform, and the strain state will not be similar. Therefore, multiple elements are needed to model single crystals to approximate the nonuniform strain state in the RVE.}. Doing so incurred additional time for generating the single crystal "basis functions", but the total time of 504s is still short compared to running the full RVE simulation at 122s each. The time required to generate the single crystal responses amounts to about 1\% of the time to generate the 400 RVE simulations. Transfer learning of SC-DeepONet was attempted with a different number of new data, and results are shown in \fref{al_shear1}. Performance metrics and predicted stress-strain curves for the model fine-tuned with 20 new data points are shown in \tref{metrics_shear} and \fref{al_shear}, respectively. The transfer learning process took 46s and the prediction time was $8\times10^{-3}$s per case, 15248 times faster than the FE simulation.
\begin{table}[h!]
    \caption{Performance metrics for the aluminum under 2\% shear}
    \centering
    \begin{tabular}{cccccc}
     Model & \vline & Rel. err. (before) [\%] & Rel. err. (after) [\%] & Mean abs. err. [MPa] & $R^2$ \\
    \hline
    SC-DeepONet & \vline  & 23.73 & 1.81 & 5.50 & 0.994 \\
    \end{tabular}
    \label{metrics_shear}
\end{table}

In this section, we only tested the SC-DeepONet and did not use the MP-DeepONet. MP-DeepONet cannot be fine-tuned via transfer learning (without modifying the branch FNN architecture) for this case as it does not have the means to distinguish between tensile (what it was initially trained on) and shear loading: the type of boundary condition is challenging to encode due to its discreteness and therefore was not included as input to MP-DeepONet (only the tensile strain magnitude was included). From this perspective, the single crystal stress-strain curves under identical boundary conditions provide a natural and mechanically meaningful method to implicitly represent only the \emph{effects} of the boundary conditions rather than trying to represent and encode the boundary conditions themselves. After all, when predicting the homogenized RVE responses, we are only interested in understanding how changing the boundary conditions affects the stress-strain curves. As seen in \tref{metrics_shear}, the SC-DeepONet trained on aluminum tensile data did not provide good prediction in shear before transfer learning, having a significant relative error of nearly 26\%. However, the mean error reduced drastically to below 3\% after exposure to only 5 new data points and continued to improve with more training data added. This again speaks to the data efficiency of the proposed architecture. From the scatter plot of predicted stress points, we see that the prediction correlated well with the CP simulation with a $R^2$ value of 0.994, and 95\% of the predictions had a relative error $\le 5\%$. From \fref{al_shear}, we see that the SC-DeepONet captures the shape of the global stress-strain curves thanks to the help of the single crystal responses and accurately predicts the elastic slope and hardening behavior. For the cases with relatively large prediction errors, it is observed that SC-DeepONet tends to under-predict the flow stress.

\subsection{Generalization to cyclic loading}
\label{sec:cyclic}
\begin{figure}[h!] 
    \centering
     \subfloat[]{
         \includegraphics[trim={0cm 0cm 0cm 0cm},clip,width=0.31\textwidth]{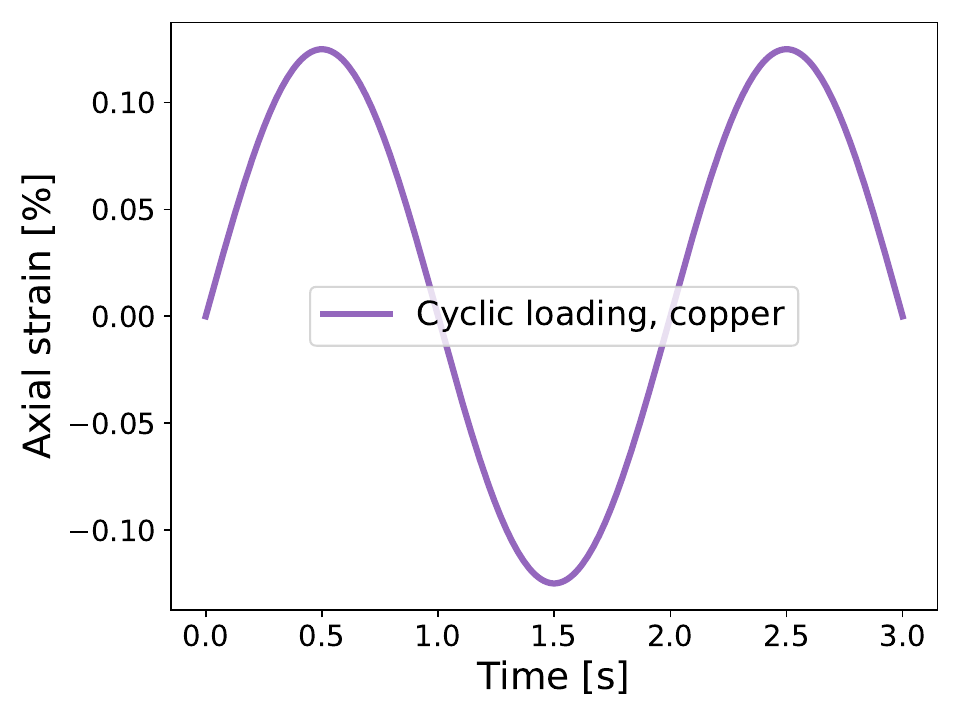}
         \label{cyc_bc}
     }
     \subfloat[]{
         \includegraphics[trim={0cm 0cm 0cm 0cm},clip,width=0.31\textwidth]{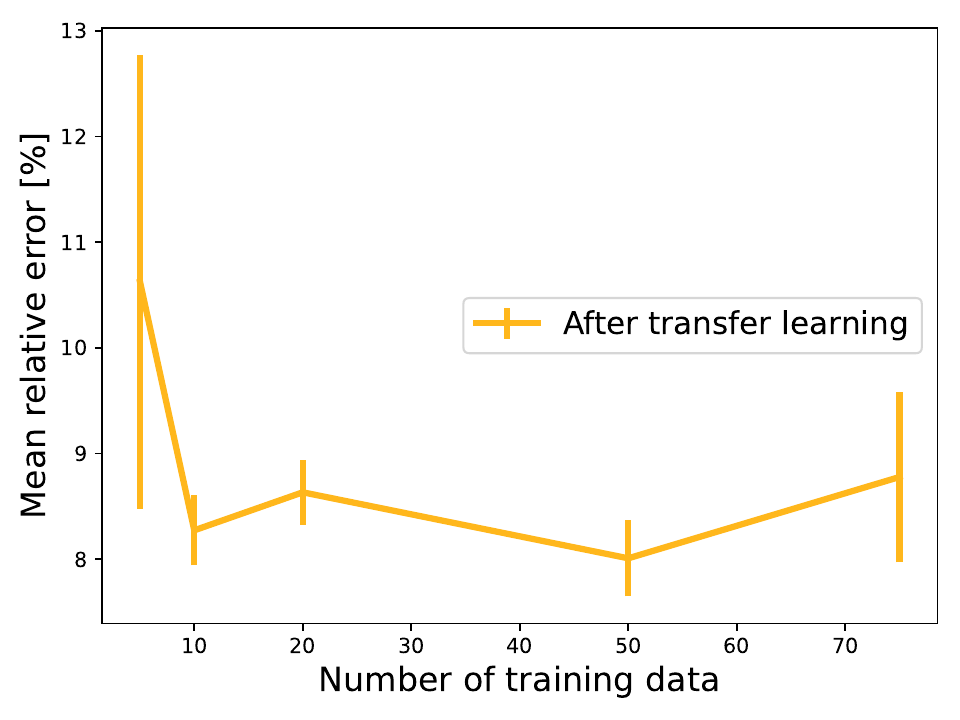}
         \label{tr_cyc}
     }
     \subfloat[]{
         \includegraphics[trim={0cm 0cm 0cm 0cm},clip,width=0.31\textwidth]{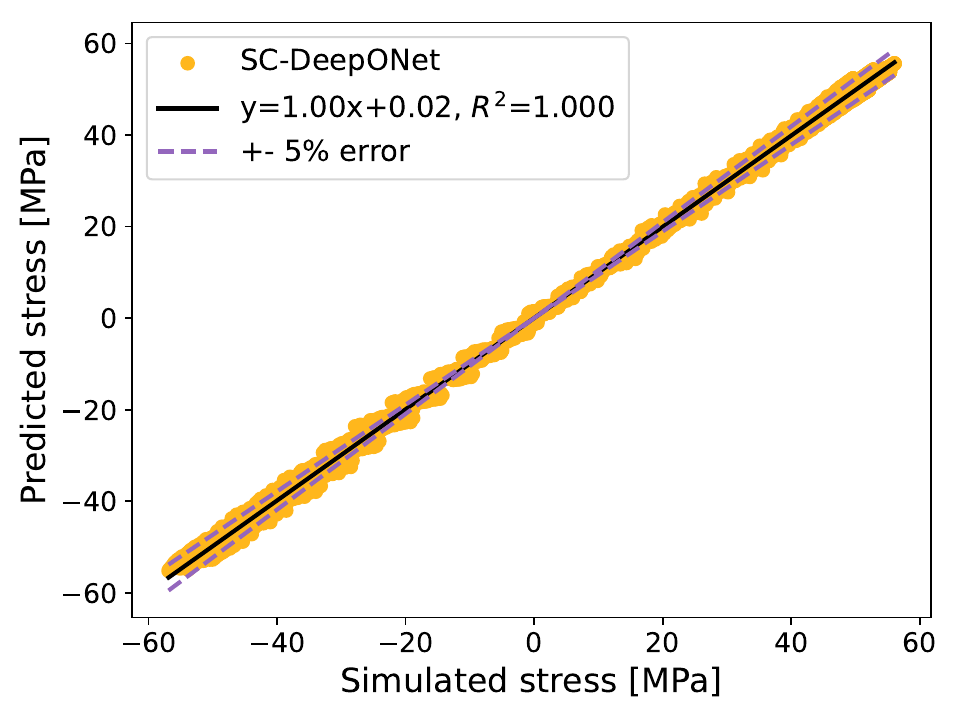}
         \label{tr_cyc_corr}
     }
    \caption{\psubref{cyc_bc} Applied axial strain as a function of time. \psubref{tr_cyc} Final SC-DeepONet prediction error for different numbers of transfer learning data points. \psubref{tr_cyc_corr} Correlation of predicted and FE simulated stress data.}
    \label{cu_cyc}
\end{figure}

\begin{figure}[h!]
\newcommand\x{0.23}
    \centering
    \begin{tabular}{ c c c c c }
    \begin{minipage}[c]{\x\textwidth}
       \centering 
        \subfloat[SC-DeepONet, best]{\includegraphics[trim={0cm 0cm 0cm 0cm},clip,width=\textwidth]{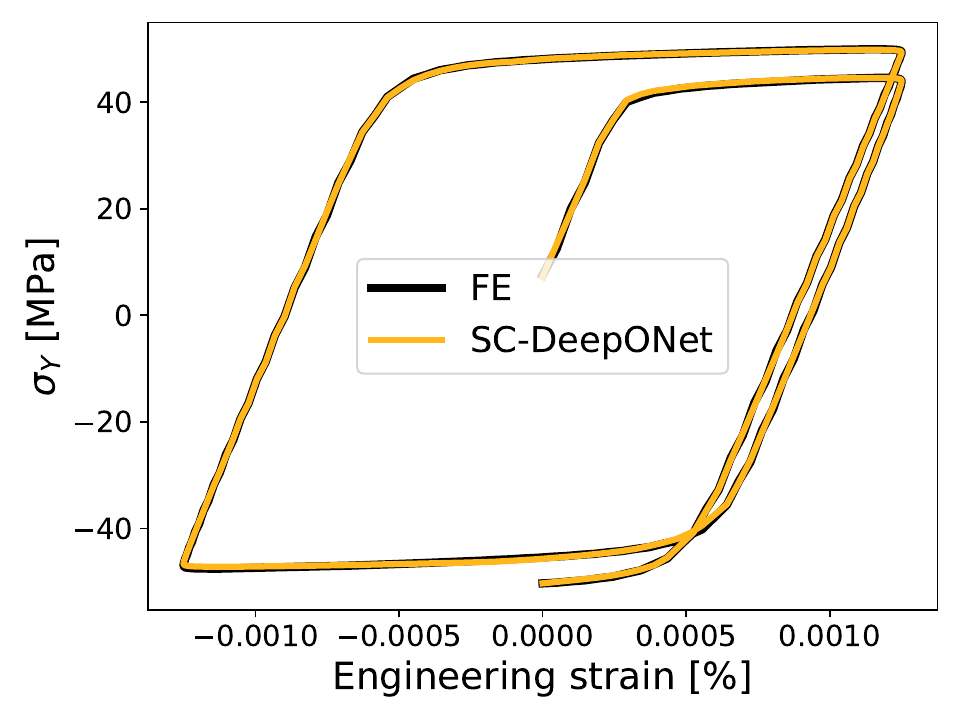}
        \label{p61}}
    \end{minipage} &
    \begin{minipage}[c]{\x\textwidth}
       \centering 
        \subfloat[SC-DeepONet, 50$^{th}$ pct]{\includegraphics[trim={0cm 0cm 0cm 0cm},clip,width=\textwidth]{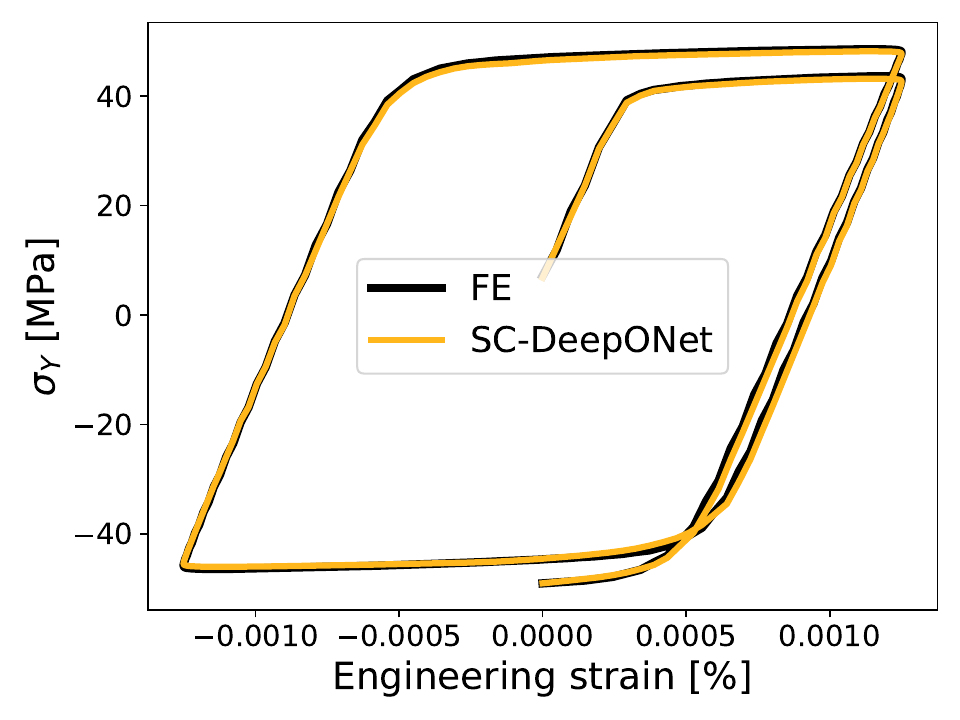}
        \label{p62}}
    \end{minipage} &
    \begin{minipage}[c]{\x\textwidth}
       \centering 
        \subfloat[SC-DeepONet, 85$^{th}$ pct]{\includegraphics[trim={0cm 0cm 0cm 0cm},clip,width=\textwidth]{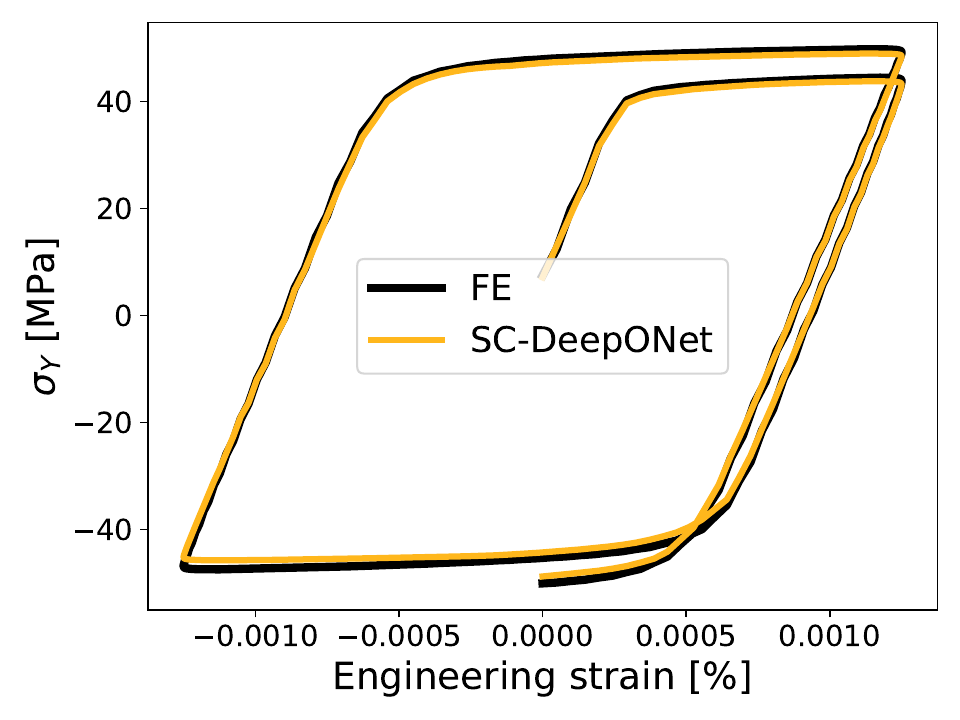}
        \label{p63}}
    \end{minipage} &
    \begin{minipage}[c]{\x\textwidth}
       \centering 
        \subfloat[SC-DeepONet, worst]{\includegraphics[trim={0cm 0cm 0cm 0cm},clip,width=\textwidth]{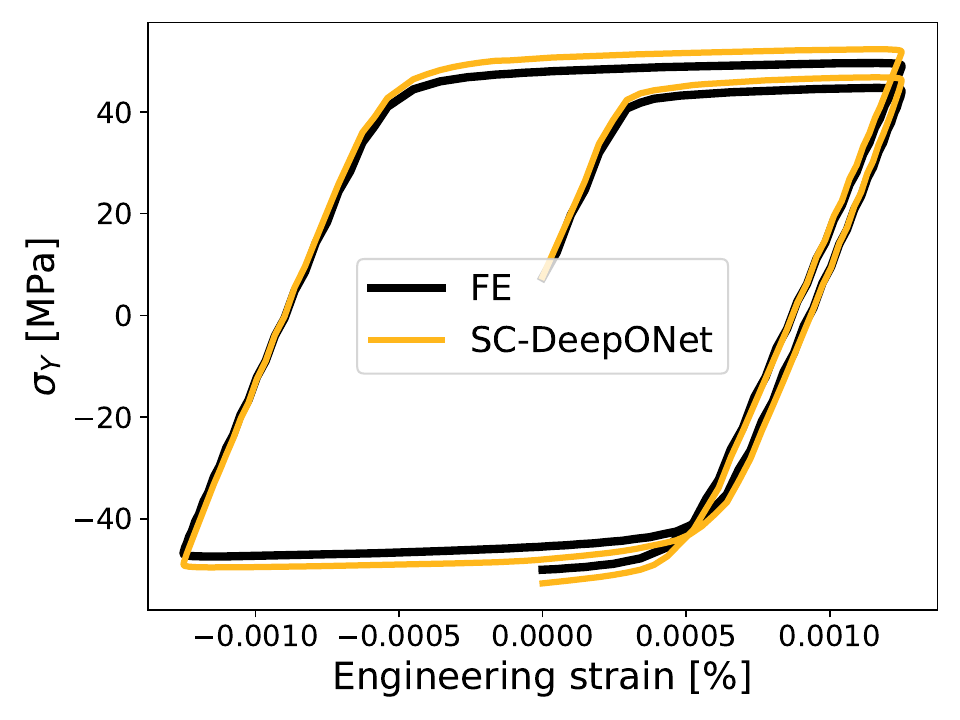}
        \label{p64}}
    \end{minipage} \\

    \end{tabular}
    \caption{Testing results for SC-DeepONet at different error percentiles after transfer learning with 20 data points, for copper under cyclic loading.}
    \label{cu_cyc2}
\end{figure}

In the last example, we investigate how a DeepONet trained on monotonic loading data of aluminum can be extended to predict the cyclic loading behavior of copper with the help of the corresponding single crystal responses. 186 RVE simulations were generated, and the average run time for each CP simulation was 2256.8s. Generating the single crystal responses (modeled by a single finite element) took 10s, significantly shorter than running a full RVE simulation. A sinusoidal, fully reversible displacement was applied in the Y-direction for a maximum engineering strain of 0.125\% (see \fref{cyc_bc}). The mean axial stress $\sigma_y$ of the RVE was recorded at 240 uniformly spaced time steps (while all previous examples only had 50 time steps). Transfer learning results with different new data points are shown in \fref{tr_cyc}. Performance metrics and predicted stress-strain curves for the model fine-tuned with 20 new data points are shown in \tref{metrics_cyclic} and \fref{cu_cyc2}, respectively. The transfer learning process took 47s, and the DeepONet prediction was over $2.8\times10^5$ times faster than the FE simulation.
\begin{table}[h!]
    \caption{Performance metrics for the copper under cyclic loading}
    \centering
    \begin{tabular}{cccccc}
     Model & \vline & Rel. err. (before) [\%] & Rel. err. (after) [\%] & Mean abs. err. [MPa] & $R^2$ \\
    \hline
    SC-DeepONet & \vline  & 154.90 & 7.34 & 0.54 & 0.999 \\
    \end{tabular}
    \label{metrics_cyclic}
\end{table}

Similar to \sref{sec:shear}, only SC-DeepONet was tested but not the MP-DeepONet. Two reasons for that are: (1) in MP-DeepONet, only the magnitude of the applied strain is used as input to encode the loading, this works well for monotonic loading but does not capture the time-dependent applied displacement in a fully reversible cyclic loading; (2) MP-DeepONet was trained to output 50 time steps only, its branch FNN has a fixed output shape of $[50\times HD]$. Therefore, it is not possible to change the number of output time steps without adding new layers to the branch FNN of MP-DeepONet (while in the current transfer learning scheme, we are only fine-tuning the weights and bias of the FNN without changing its architecture). Neither of those is a limitation of the SC-DeepONet using single crystal responses as inputs. First, the single crystal stress-strain curves carry information on the time-dependent axial displacement, removing the need for explicit information on the applied boundary conditions. Secondly, the number of output time steps for the SC-DeepONet follows directly from the input functions, as each time step is treated as an independent sample with 36 features (stress from the 36 single crystal samples at that time). Therefore, no changes to the branch FNN architecture are warranted when predicting stress-strain curves with different temporal resolutions. These two observations show the versatility of the SC-DeepONet compared to using material properties and loading conditions as direct inputs to the MP-DeepONet. From \tref{metrics_cyclic}, we see that the SC-DeepONet before any transfer learning lead to significant prediction errors, but this is expected due to the profound differences in the RVE responses (comparing \fref{c1} with \fref{c4}) caused by a change of both material properties and loading conditions. With only 10 new data points used in transfer learning, the relative error dropped to below 10\%. It is also worth pointing out that the use of relative error might be biased in this case since the axial stress can become 0 during cyclic loading, and relative error for a small stress quantity close to 0 can be quite large. Nonetheless, with 20 data points in transfer learning, 96.9\% of the predicted stresses had a relative error below 5\%, and the mean absolute stress error was only around 0.5 MPa. Different stress-strain curves in \fref{cu_cyc2} also show satisfactory agreement with the corresponding FE simulation results, indicating that the SC-DeepONet can effectively capture the cyclic response of polycrystal RVEs given information on the single crystal responses.

\section{Conclusions, limitations, and future work}
\label{sec:conc}
The present work introduces a novel material response-informed DeepONet architecture to predict the mean-field stress-strain responses for polycrystalline RVEs. It leverages a ResUNet in the trunk (similar to that of \cite{HE2023116277}) to encode the grain orientations in the polycrystal RVE and learn the interactions of a voxel in the domain with its neighbors. The most novel contribution of the current work is the use of single crystal stress-strain responses as inputs to the branch network of the DeepONet. The single crystal response curves contain information on the material properties, boundary conditions, and loading, effectively serving as "basis functions" to reconstruct the RVE response. A dot product operation combines the encoded output from the branch and trunk networks to produce the prediction. In this view, the SC-DeepONet combines the information on grain-to-grain interaction (extracted from the ResUNet trunk network) and information on the individual grain behavior (inputs to the FNN branch network) to produce a nonlinear, microstructure-dependent rule of mixture as an estimation of the RVE response. It captures all key elements that influence the global response in crystal plasticity simulations. The proposed DeepONet architecture is compared to a more traditional approach, where the material properties and applied strain values are used as direct inputs to the DeepONet branch network. When trained on data involving a single material and loading, both DeepONets can deliver satisfactory predictions of global stress-strain curves. However, when asked to generalize to different material and tensile strain levels, the MP-DeepONet that takes material properties as inputs suffers from large prediction errors and cannot capture the more anisotropic elastic behavior in the new data. On the other hand, SC-DeepONet gives less than 3\% relative error even before any transfer learning and reduces to about 1\% with only 20 new data points. SC-DeepONet accurately predicts the RVE stress-strain curve up to about 85$^{th}$ percentile error and captures the correct elastic and hardening behavior even in the worst case. SC- DeepONet is further tested on different loading conditions, such as shear and cyclic tensile loading. The SC-DeepONet can be fine-tuned to produce a mean relative error of 1.8\% and 7.3\%, respectively, with only 20 new data points. The predicted stress-strain curves in both cases agree well with the CP FE simulations. Training of the SC-DeepONet is efficient, the baseline training on 800 data points took less than 5mins, while all transfer learning procedures finished within 1min. Once fine-tuned, the prediction speed is nearly 10000 times faster than FE simulations. Generating the single crystal responses is also computationally cheap, taking less than 10s for single element cases and about 1\% of the RVE data generation time if the single crystal has to be modeled with multiple finite elements. It is therefore concluded that the proposed SC-DeepONet is highly data efficient. It overcomes the limitations of using material properties in the branch network, such as being unable to encode different types of boundary conditions, and not being able to predict different number of time steps without modifying the branch network architecture. These results are evidence that the new architecture is versatile and highly generalizable.

The current DeepONet can be a valuable tool in efficient multi-scale analyses. Due to the high computational expense, it is often prohibitive to run CP simulations on engineering structures. Yet the ability of CP simulations to capture the effect of local microstructure on mechanical response can be vital for predicting the performance of additively manufactured components, where the local microstructure can vary significantly due to print conditions and geometry. In this sense, it is sufficient to take a homogenization approach, where the microstructure-informed global response of a small representative volume is calculated from CP, and continuum plasticity models can be used to capture the location-dependent elastic-plastic behavior. This process inherently demands many CP simulations on RVEs with different microstructures and can change if the part is manufactured with a different material. The current SC-DeepONet provides an efficient surrogate model for evaluating the mean-field responses given a new microstructure, material, and loading condition. Although not mentioned in the current work, the SC-DeepONet can be easily extended to treat 3D polycrystal RVEs using a 3D analog of the ResUNet and more single-crystal responses that uniformly sample the full grain orientation space. Once properly trained, the SC-DeepONet can efficiently generate location-dependent homogenized elastic-plastic material properties for additively manufactured components, where the underlying microstructure can come from part-scale process simulations. It furnishes a seamless and efficient integrated computational materials engineering (ICME) workflow \citep{panchal2013key,liu2018vision}, and the differentiable NN is well-suited for gradient-based optimization to optimize mechanical properties. 

Finally, it is worth reiterating the assumptions and limitations of the current work. The central idea of this work is to use the single crystal responses under identical boundary conditions as input "basis functions" and approximate the RVE response by constructing a nonlinear rule of mixture. It requires the assumption that the mean strain states between the single crystal samples and the polycrystal RVE are similar, since strain is not directly passed into the SC-DeepONet. It is emphasized that this is only valid in an average sense, and the local strain state at individual finite elements within the RVE might be drastically different from that experienced by the single crystals. Therefore, the current SC-DeepONet architecture cannot be used to predict the full contour of the stress field since strain similarity does not hold on a local level. It should only be used in a homogenization setting where the local response is unimportant and only the mean-field response is of interest. In future work, we will extend the proposed DeepONet to account for 3D polycrystal RVEs and expose the network to more realistic microstructures generated using cellular automata simulation of additive manufacturing.

\section*{Replication of results}
The data and source code that support the findings of this study can be found at \url{https://github.com/Jasiuk-Research-Group}. \textcolor{red}{Note to editor and reviewers: the link above will be made public upon the publication of this manuscript. During the review period, the data and source code can be made available upon request to the corresponding author.}

\section*{Conflict of interest}
The authors declare that they have no conflict of interest.

\section*{Acknowledgements}
The authors would like to thank the National Center for Supercomputing Applications (NCSA) at the University of Illinois, and particularly its Research Consulting Directorate, the Industry Program, and the Center for Artificial Intelligence Innovation (CAII) for their support and hardware resources. This research is a part of the Delta research computing project, which is supported by the National Science Foundation (award OCI 2005572) and the State of Illinois, as well as the Illinois Computes program supported by the University of Illinois Urbana-Champaign and the University of Illinois System. Finally, the authors would like to thank Professors George Karniadakis, Lu Lu, and the Crunch team at Brown, whose original work with DeepONets inspired this research.

\section*{CRediT author contributions}
\textbf{Junyan He}: Methodology, Formal analysis, Investigation, Writing - Original Draft.
\textbf{Deepankar Pal}: Supervision, Writing - Review \& Editing.
\textbf{Ali Najafi}: Supervision, Writing - Review \& Editing.
\textbf{Diab Abueidda}: Supervision, Writing - Review \& Editing.
\textbf{Seid Koric}: Supervision, Resources, Writing - Review \& Editing, Funding Acquisition.
\textbf{Iwona Jasiuk}: Supervision, Writing - Review \& Editing.

\bibliographystyle{unsrtnat}
\setlength{\bibsep}{0.0pt}
{\scriptsize \bibliography{References.bib} }
\end{document}